\documentclass[twocolumn,prd,aps,showpacs,nofootinbib]{revtex4}
\usepackage{graphicx}
\usepackage{dcolumn}
\usepackage{bm}
\usepackage{epsfig}
\usepackage{amssymb}
\usepackage{amsmath}

\def\fun#1#2{\lower3.6pt\vbox{\baselineskip0pt\lineskip.9pt
   \ialign{$\mathsurround=0pt#1\hfil##\hfil$\crcr#2\crcr\sim\crcr}}}

 \def\tskip{\setlength{\tskip}{5pt}}
\def\colwidth{\setlength{\colwidth}{3.5in}}

\newcommand{\lsim}{\mathrel{\hbox{\rlap{\lower.55ex\hbox{$\sim$}} \kern-.3em \raise.4ex \hbox{$<$}}}}
\newcommand{\gsim}{\mathrel{\hbox{\rlap{\lower.55ex\hbox{$\sim$}} \kern-.3em \raise.4ex \hbox{$>$}}}}
\newcommand{\intderiv}{{\mbox{ }\rm d}}
\newcommand{\deriv}{{\rm d}}
\newcommand{\vesc}{v_{\rm esc}}
\newcommand{\vcirc}{v_{\rm c}}
\newcommand{\scoh}{\sigma_\mathrm{coh}}
\newcommand{\sn}{\sigma_\mathrm{Dn}}
\newcommand{\stot}{\sigma_\mathrm{tot}}
\newcommand{\mred}{m_\mathrm{red}}
\newcommand{\mdm}{m_\mathrm{dm}}
\newcommand{\mnuc}{m_\mathrm{T}}

\begin{document}

\title{Constraints on the Interactions between Dark Matter and Baryons from the X-ray Quantum Calorimetry Experiment}
\author{Adrienne L. Erickcek$^{1,2}$, Paul J. Steinhardt$^{2,3}$, Dan McCammon$^4$, and Patrick C. McGuire$^5$}
\affiliation{$^1$Division of Physics, Mathematics, \& Astronomy, California Institute of Technology, Mail Code 103-33, Pasadena, CA 91125, USA}
\affiliation{$^2$Department of Physics, Princeton University, Princeton, NJ 08544, USA}
\affiliation{$^3$Princeton Center for Theoretical Physics, Princeton University, Princeton, NJ 08544, USA}
\affiliation{$^4$Department of Physics, University of Wisconsin, Madison, WI 53706, USA}
\affiliation{$^5$McDonnell Center for the Space Sciences, Washington University, St. Louis, M0 63130, USA}
\pacs{95.35.+d, 12.60.-i, 29.40.Vj}
\begin{abstract}
Although the rocket-based X-ray Quantum Calorimetry (XQC) experiment was designed for X-ray spectroscopy, the minimal shielding of its calorimeters, its low atmospheric overburden, and its low-threshold detectors make it among the most sensitive instruments for detecting or constraining strong interactions between dark matter particles and baryons. We use Monte Carlo simulations to obtain the precise limits the XQC experiment places on spin-independent interactions between dark matter and baryons, improving upon earlier analytical estimates.  We find that the XQC experiment rules out a wide range of nucleon-scattering cross sections centered around one barn for dark matter particles with masses between 0.01 and $10^5$~GeV.  Our analysis also provides new constraints on cases where only a fraction of the dark matter strongly interacts with baryons.
\end{abstract}
\maketitle
\section{INTRODUCTION}
From Vera Rubin's discovery that the rotation curves of galaxies remain level to radii much greater than predicted by Keplerian dynamics \cite{Rot} to the Wilkinson Microwave Anisotropy Probe (WMAP) measurement of the cosmic microwave background (CMB) temperature anisotropy power spectrum \cite{WMAP3}, observations indicate that the luminous matter we see is only a fraction of the mass in the Universe.   The three-year WMAP CMB anisotropy spectrum is best-fit by a cosmological model with $\Omega_{\mathrm m} = 0.241 \pm 0.034$ and a baryon density that is less than one fifth of the total mass density.  The cold collisionless dark matter (CCDM) model has emerged as the predominant paradigm for discussing the missing mass problem.  The dark matter is assumed to consist of non-relativistic, non-baryonic, weakly interacting particles, often referred to as Weakly Interacting Massive Particles (WIMPs).  

Although the CCDM model successfully predicts observed features of large-scale structure at scales greater than one megaparsec \cite{Bahcall}, there are indications that it may fail to match observations on smaller scales.  Numerical simulations of CCDM halos \cite{NFW, Kravtsov, Moore, Ghigna, Power03, Navarro04, Hayashi04, Diemand04, Diemand05} imply that CCDM halos have a density profile that increases sharply at small radii ($\rho \sim r^{-1.2}$ according to Ref.~\cite{Diemand05}).  These predictions conflict with lensing observations of clusters \cite{Tyson, Sand04} that indicate the presence of constant-density cores.  X-ray observations of clusters have found cores in some clusters, although density cusps have also been observed \cite{KH04, PAP05, VF06}.  On smaller scales, observations of dwarf and low-surface-brightness galaxies \cite{Flores, Moore94, Burk, deBlok, MB98, BMR01, Marchesini02} indicate that these dark matter halos have constant-density cores with lower densities than predicted by numerical simulations.  Observations also indicate that cores are predominant in spiral galaxies as well, including the Milky Way \cite{BE01, Salucci01, SBLBG05}.  Numerical simulations of CCDM halos also predict more satellite halos than are observed in the Local Group \cite{GSSMoore, KKVP99} and fossil groups \cite{DL04}.  

Astrophysical explanations for the discord between the density profiles predicted by CCDM simulations and observations have been proposed: for instance, dynamical friction may transform density cusps into cores in the inner regions of clusters \cite{ZHPC04}, and the triaxiality of galactic halos may mask the true nature of their inner density profiles \cite{HN06}.  There are also models of substructure formation that explain the observed paucity of satellite halos \cite{BKW00, Benson02, KGK04, Moore06}.  

Another possible explanation for the apparent failure of the CCDM model to describe the observed features of dark matter halos is that dark matter particles scatter strongly off one another.  The discrepancies between observations and the CCDM model are alleviated if one introduces a dark matter self-interaction that is comparable in strength to the interaction cross section between neutral baryons \cite{oriSIDM, Wandelt}:
\begin{equation}
\frac{\sigma_\mathrm{DD}}{\mdm} = 8 \times 10^{-25} - 1 \times 10^{-23} \mbox{cm$^2$ GeV$^{-1}$},
\end{equation}
where $\sigma_\mathrm{DD}$ is the cross section for scattering between dark matter particles and $\mdm$ is the mass of the dark matter particle.  Numerical simulations have shown that introducing dark matter self-interactions within this range reduces the central slope of the halo density profile and reduces the central densities of halo cores, in addition to destroying the extra substructure \cite{Dave, AS05}.  

The numerical coincidence between this dark matter self-interaction cross section and the known strong-interaction cross section for neutron-neutron or neutron-proton scattering has reinvigorated interest in the possibility that dark matter interacts with itself and with baryons through the strong nuclear force. We refer to dark matter of this type as ``strongly interacting dark matter'' where ``strong" refers specifically to the strong nuclear force.  Strongly interacting dark matter candidates include the dibaryon \cite{Farrar03, FZ06}, the Q-ball \cite{KS01}, and O-helium \cite{Khlopov}. 

Surprisingly, the possibility that the dark matter may be strongly interacting is not ruled out.  While there are numerous experiments searching for WIMPs, they are largely insensitive to dark matter that interacts strongly with baryons.  The reason is that WIMP searches are typically conducted at or below ground level based on the fact that WIMPs can easily penetrate the atmosphere or the Earth, whereas strongly interacting dark matter is multiply scattered and thermalized by the time it reaches ground level and its thermal kinetic energy is too small to produce detectable collisions with baryons in WIMP detectors.  Consequently, there are few experiments capable of detecting strongly interacting dark matter directly.  Starkman {\it et al}. \cite{Stark} summarized the constraints on strongly interacting dark matter from experiments prior to 1990, and these constraints were later refined \cite{McGuire94, McGuire95, Wandelt, MS01}.  The strength of dark matter interactions with baryons may also be constrained by galactic dynamics \cite{Stark}, cosmic rays \cite{Stark, CFPW02}, Big Bang nucleosynthesis (BBN) \cite{CFPW02}, the CMB \cite{CHS02}, and large-scale structure \cite{CHS02}.

The X-ray Quantum Calorimetry (XQC) project launched a rocket-mounted micro-calorimeter array in 1999 \cite{big}.  At altitudes above 165 km, the XQC detector collected data for a little less than two minutes.  Although its primary purpose was X-ray spectroscopy, the limited amount of shielding in front of the calorimeters and the low atmospheric overburden makes the XQC experiment a sensitive detector of strongly interacting dark matter. 

In this article, we present a new numerical analysis of the constraints on spin-independent interactions between dark matter particles and baryons from the XQC experiment using Monte Carlo simulations of dark matter particles interacting with the XQC detector and the atmosphere above it.  Our work is a significant improvement upon the earlier analytic estimates presented by some of us in Refs.~\cite{Wandelt, MS01} because it accurately models the dark matter particle's interactions with the atmosphere and the XQC instrument.  Our calculation here also supersedes the analytic estimate by Zaharijas and Farrar \cite{ZF05} because they only considered a small portion of the XQC data and did not include multiple scattering events nor the overburden of the XQC detector.  We restrict our analysis to spin-independent interactions because the XQC calorimeters are not highly sensitive to spin-dependent interactions.  Only a small fraction of the target nuclei in the calorimeters have non-zero spin; consequently, the bound on spin-dependent interactions between baryons and dark matter from the XQC experiment is about four orders of magnitude weaker than the bound on spin-independent interactions \cite{ZF05}.

This article is organized as follows.  In Section \ref{sec:XQC} we summarize the specifications of the XQC detector.  We then review dark matter detection theory in Section \ref{sec:darkmatterdetect}.  This Section includes a discussion of coherent versus incoherent scattering and how we account for the loss of coherence in our analysis.  A complete description of our analysis follows in \ref{sec:analysis}, and our results are presented in Section \ref{sec:results}.  Finally, in Section \ref{sec:conclusion}, we summarize our findings and compare the constraints to strongly interacting dark matter from the XQC experiment to those from other experiments.

\section{THE XQC EXPERIMENT}
\label{sec:XQC}
Calorimetry is the use of temperature deviations to measure changes in the internal energy of a material. By drastically reducing the specific heat of the absorbing material, the use of cryogenics in calorimetry allows the absorbing object to have a macroscopic volume and still be sensitive to minute changes in energy.  These detectors are sensitive enough to register the energy deposited by a single photon or particle and gave birth to the technique of ``quantum calorimetry,'' the thermal measurement of energy quanta.  

The quantum calorimetry experiment \cite{big} we use to constrain interactions between dark matter particles and baryons is the second rocket-born experiment in the XQC (X-ray Quantum Calorimetry) Project, a joint undertaking of the University of Wisconsin and the Goddard Space Flight Center \cite{rocket, microcal}.  It launched on March 28, 1999 and collected about 100 seconds of data at altitudes between 165 and 225 km above the Earth's surface.  The detector consisted of thirty-four quantum calorimeters operating at a temperature of 0.06 K; for detailed information on the XQC detector functions, please refer to Refs.~\cite{big, microcal}.  These detectors were separated from the exterior of the rocket by five thin filter panes \cite{big}.  The small atmospheric overburden at this altitude and the minimal amount of shielding in front of the calorimeters makes this experiment a promising probe of strongly interacting dark matter.  

\begin{figure}
\centerline{\epsfig{file=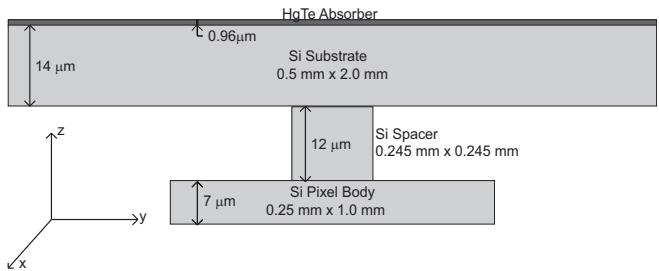, width=3.4in}}
\caption{A vertical cross section of an XQC calorimeter.  The relative thicknesses of the layers are drawn to scale, as are their relative lengths, but the two scales are not the same.  To facilitate the display of the layers, the vertical dimension has been stretched relative to the horizontal dimension.}
\label{fig:detside}
\end{figure}

\begin{figure}
\centerline{\epsfig{file=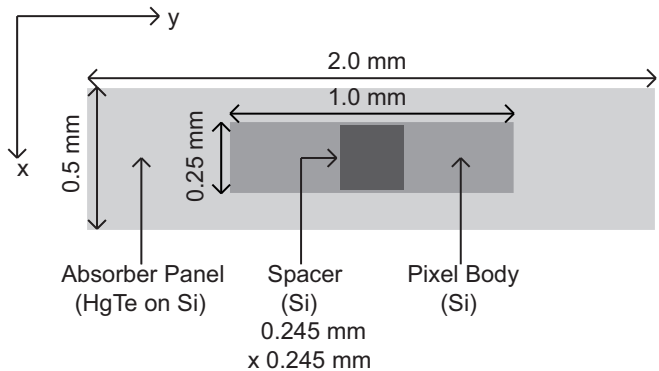, width=3.4in}}
\caption{A top view of an XQC calorimeter.  The absorber is the top layer and underneath it lies the spacer, followed by the pixel body.  These dimensions are drawn to scale.}
\label{fig:detabove}
\end{figure}

The absorbers in the XQC calorimeters are composed of a thin film of HgTe (0.96 $\mu$m thick) deposited on a silicon (Si) substrate that is 14 $\mu$m thick.  The absorbers rest on silicon spacers and silicon pixel bodies.  Figs.~\ref{fig:detside} and \ref{fig:detabove} show side and top views of the detectors with the dimensions of each layer.  Temperature changes in all four components are measured by the calorimeter's internal thermometer.  The calorimeters report the average temperature over an integration time of 7 ms in order to reduce the effect of random temperature fluctuations on the measurement.  Multiple scatterings by a dark matter particle will register as a single event because the time it takes the dark matter particle to make its way through the calorimeter is small compared to the integration time.

The detector array consists of two rows of detectors, with seventeen active calorimeters and one inactive calorimeter in each row, and is located at the bottom of a conical detector chamber.   
Within a 32-degree angle from the detector normal, the incoming particles only pass through the aforementioned filters.  The five filters are located 2 mm, 6 mm, 9 mm, 11 mm and 28 mm above the detectors.  Each filter consists of a thin layer of aluminum (150 \AA) supported on a parylene (CH) substrate (1380 \AA).  The pressure inside the chamber is less than $10^{-6}$ Torr.  At this level of evacuation, a dark matter particle with a mass of $10^6$ GeV and a baryon interaction cross section of $10^6$ barns, would have less than a 20\% chance of colliding with an air atom in the chamber.  Therefore, we assume that the chamber is a perfect vacuum in our analysis.   

While the atmospheric pressure at the altitudes at which the detector operated is about $10^{-8}$ times the atmospheric pressure at sea level, the atmospheric overburden of the XQC detector is still sufficient to scatter incoming strongly interacting dark matter particles.  Simulating a dark matter particle's path through the atmosphere requires number-density profiles for all the molecules in the atmosphere.  These profiles were obtained using the MSIS-E-90 model\footnote{Available at http://modelweb.gsfc.nasa.gov/models/msis.html} for the time (1999 March 28 9:00 UT) and location (White Sands Missile Range, New Mexico) of the XQC rocket launch.  

\begin{figure}
\centerline{\epsfig{file=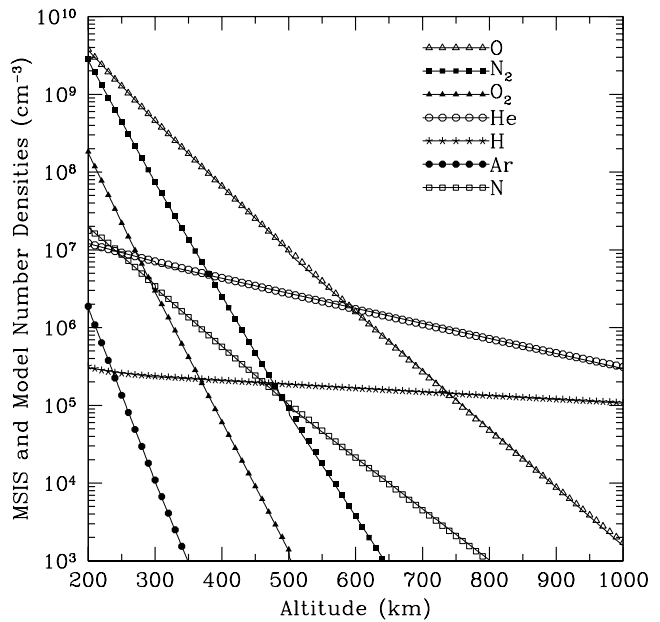, width=3.4in}}
\caption{The points depict the MSIS-E-90 density profiles for the seven most prevalent constituents of the atmosphere above the XQC detector, and the lines show the piecewise exponential fits used in our analysis.}
\label{fig:MSISmodel}
\end{figure}

During the data collection period, the average altitude of the XQC rocket was 201.747 km. At this altitude and above, the primary constituents of the atmosphere are molecular and atomic oxygen, molecular and atomic nitrogen, helium, atomic hydrogen, and argon.  The MSIS-E-90 model provides tables of the number densities of each of these seven chemical species.  In our analysis, computational efficiency demanded that we fit analytic functions to these data.  We found exponential fits for the density profiles in three altitude ranges: 200-300 km, 300-500 km and 500-1000 km.  The error in the probability of a collision between a dark matter particle and an element of the atmosphere introduced by using these fits instead of the original data is 0.02\%.  Fig.~\ref{fig:MSISmodel} shows the number density profiles provided by the MSIS-E-90 model and the exponential fits used to model the data.    

The XQC detector collected data for a total of 150 seconds.  During these 150 seconds of activity, the thirty-four individual calorimeters were not all operational at all times.   Furthermore, events that could not be accurately measured by the calorimeters and events attributed to cosmic rays hitting the base of the detector array were removed from the XQC spectrum, and these cuts also contribute to the dead time of the system.  Specifically, events that arrived too close together for the calorimeters to accurately measure distinct energies were discarded.  This criterion removed 12\% of the observed events and the resulting loss of sensitivity was included in the dead time of the calorimeters.  
When a cosmic ray penetrates the silicon base of the detector array, the resulting temperature increase is expected to register as multiple, nearly simultaneous, low-energy events on nearby calorimeters.
To remove these events from the spectrum, we cut out events that were part of either a pair of events in adjacent detectors or a trio of events in any of the detectors that arrived within 3 ms of each other and had energies less than 2.5 keV.
This procedure was expected to remove more than 97\% of the events that resulted from cosmic rays hitting the base of the array.  Nearly all of the events attributed to heating from cosmic rays had energies less than 300 eV, and a high fraction of the observed low-energy events were included in this cut.  For example, seventeen of the observed twenty-four events with energies less than 100 eV were removed.
The expected loss of sensitivity due to events being falsely attributed to cosmic rays was included in the calculated dead time of the calorimeters.  Once all the dead time is accounted for, the 150 seconds of data collection is equivalent to 100.7 seconds of observation with all thirty-four calorimeters operational.

\begin{figure}
\centerline{\epsfig{file=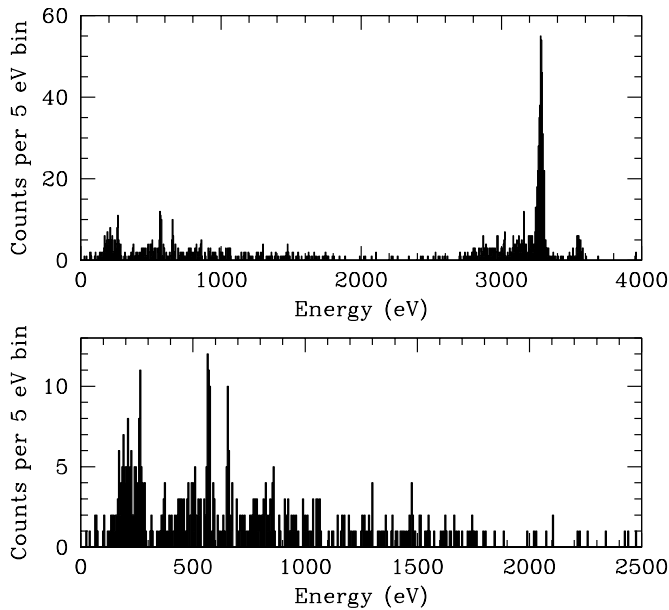, width=3.5in}}
\caption{Top panel: The XQC energy spectrum from 0 - 4 keV in 5 eV bins.  This spectrum does not have non-linearity corrections applied (see Ref.~\cite{big}), so the calibration lines at 3312 eV and 3590 eV appear slightly below their actual energies.  The cluster of counts to the left of each calibration peak result from X-rays passing through the HgTe layer and being absorbed in the Si components where up to 12\% of the energy may then be trapped in metastable states. 
Bottom panel: The XQC energy spectrum from 0 - 2.5 keV in 5 eV bins.  This spectrum, combined with the over-saturation rate of 0.6 events per second with energies greater than 4000 eV, was used in our analysis.}
\label{totalspect}
\end{figure}    

The XQC calorimeters are capable of detecting energy deposits that exceed 20 eV, but full sensitivity is not reached until the energy surpasses 36 eV, and for approximately half of the detection time, the detector's lower threshold was set to 120 eV.  The calorimeters cannot resolve energies above 4 keV, and the 2.5-4 keV spectrum is dominated by the detector's interior calibration source: a ring of 2$\mu$Ci $^{41}$Ca that generates K$\alpha$ and K$\beta$ lines at 3312 eV and 3590 eV, respectively.  We refer the reader to Ref.~\cite{big} for a complete discussion of the calibration of the detector.  These limitations restrict the useful portion of the XQC spectrum to 0.03-2.5 keV.  This spectrum is shown in Fig.~\ref{totalspect}, along with the full spectrum from 0-4 keV.  The XQC field of view was centered on a region of the sky known to have an enhanced X-ray background in the 100-300 eV range, possibly due to hot gas in the halo, and this surge in counts can be seen in Fig.~\ref{totalspect}.  In addition to the information present in this spectrum, we know that the XQC detector observed an average over-saturation event rate of 0.6 per second.  This corresponds to a total of 60 events that deposited more than 4000 eV in a calorimeter.  In Section \ref{sec:comparison}, we describe how we use the observed spectrum between 29 eV and 2500 eV and the integrated over-saturation rate to constrain the total cross section for elastic scattering between dark matter particles and nucleons.

\section{DETECTING DARK MATTER}
\label{sec:darkmatterdetect}
\subsection{Incidence of dark matter particles}
\label{sec:darkmatterwind}

The expected flux of dark matter particles into the detector depends on the density of the dark matter halo in the Solar System.  Unfortunately, the local dark matter density is unknown and the range of theoretical predictions is wide.  By constructing numerous models of our galaxy with various dark matter density profiles and halo characteristics, rejecting those models that contradict observations, and finding the distribution of local dark matter densities in the remaining viable models, Ref.~\cite{Gates} predicted that the local dark matter density is between 0.3 and 0.7 GeV cm$^{-3}$ assuming that the dark matter halo is flattened, and the predicted local density decreases as the halo is taken to be more spherical.  Another approach \cite{Moore01} used numerical simulations of galaxies similar to our own to find the dark matter density profile and then fit the profile parameters to Galactic observations, predicting a mean local dark matter density between 0.18 GeV cm$^{-3}$ and 0.30 GeV cm$^{-3}$. Given that it lies in the intersection of these two ranges, we use the standard value of 0.3 GeV cm$^{-3}$ for the local dark matter density in our primary analysis. This assumption ignores the possible presence of dark matter streams or minihalos, which do occur in numerical simulations \cite{Moore01} and could lead to local deviations from the mean dark matter density.

We also assume that the velocities of the dark matter particles with respect to the halo are isotropic and have a bounded Maxwellian distribution: the probability that a particle has a velocity within a differential volume in velocity-space centered around a given velocity $\vec{v}$ is
\begin{equation} 
P(\vec{v}) =  \left\{ \begin{array}{ll} 
                        \frac{1}{k} \exp\left[-\frac{v^2}{v_0^2}\right] \intderiv^3\vec{v}& 
			\mbox{ if $v \leq \vesc$},\\
			0& 
			 \mbox{ if  $v > \vesc$}.
                        \end{array}  \right.
\label{maxwell}
\end{equation}
where $v_0$ is the dispersion velocity of the halo, $\vesc$ is the Galactic escape velocity at the Sun's position, and $k$ is a normalization factor \cite{Lewin}:
\begin{equation} 
k =  (\pi v_0^2)^{3/2} \left[\mathrm{erf}\left(\frac{\vesc}{v_0}\right) - \left(\frac{2}{\sqrt{\pi}}\right) \frac{\vesc}{v_0} e^{-\frac{\vesc^2}{v_0^2}}\right].
\label{kfin}
\end{equation}
Numerical simulations indicate that dark matter particle velocities may not have an isotropic Maxwellian distribution \cite{Moore01}.  Ref.~\cite{Green03} examines how assuming a more complicated velocity distribution would alter the flux of dark matter particles into an Earth-based detector.

Given the flat rotation curve of the spiral disk at the Sun's radius and beyond and assuming a spherical halo, the local dispersion speed $v_0$ is the maximum rotational velocity of the Galaxy $\vcirc$ \cite{Drukier}.  Reported values for the rotational speed include $222\pm 20$ km s$^{-1}$ \cite{v1}, $228 \pm 19$ km s$^{-1}$ \cite{v2}, $184 \pm 8$ km s$^{-1}$ \cite{newconstants} and $230 \pm 30$ km s$^{-1}$ \cite{Kochanek96}.  Recent measurements of the Galaxy's angular velocity have yielded values of $\Omega_\mathrm{gal} = 28 \pm 2$ km s$^{-1}$ kpc$^{-1}$ \cite{v3} and $32.8 \pm 2$ km s$^{-1}$ kpc$^{-1}$ \cite{v4}.  If the Sun is located 8.0 kpc from the Galactic center, these angular velocities correspond to tangential velocities $224 \pm 16$ km s$^{-1}$ and $262\pm16$ km s$^{-1}$ respectively.  We adopt $\vcirc = 220 \pm 30$ km s$^{-1}$ as a centrally conservative value for the Galaxy's circular velocity at the Sun's location.

The final parameter we need to obtain the dark matter's velocity distribution is the escape velocity in the Solar vicinity.  The largest observed stellar velocity at the Sun's radius in the Milky Way is 475 km s$^{-1}$, which establishes a lower bound for the local escape velocity \cite{starvel}.  Ref.~\cite{LT90} used the radial motion of Carney-Latham stars to determine that the escape velocity is between 450 and 650 km s$^{-1}$ to 90\% confidence, and Ref.~\cite{Rave06} obtained a 90\% confidence interval of 498 to 608 km s$^{-1}$ from observations of high-velocity stars.   A kinematic derivation of the escape velocity \cite{Drukier} gives
\begin{equation}
\vesc^2 = 2v_c^2\left[1+\ln\left(\frac{R_\mathrm{gal}}{R_0}\right)\right],
\label{flatvesc}
\end{equation}
where $R_0$ is the distance from the Sun to the center of the Galaxy, and $R_\mathrm{gal}$ is radius of the Galaxy.  Observations of other galaxies suggest that our galaxy extends to about 100 kpc \cite{Drukier}, and observations of Galactic satellites indicate that the Galaxy's flat rotation curve extends to at least 110 kpc \cite{Kochanek96}.  The commonly accepted value for the Solar radius is $R_0 = 8.0$ kpc \cite{solarrad}.  Recent measurements include $R_0 = 7.9\pm 0.3$ kpc \cite{solarrad2} and $R_0 = 8.01 \pm 0.44$ kpc \cite{Avedisova05}, and a compilation of measurements over the past decade \cite{Avedisova05} yields an average value of $R_0 = 7.80 \pm 0.33$ kpc.  To estimate the escape velocity, we use 100 kpc as a conservative estimate of the Galactic radius and the standard value $R_0 = 8.0$ kpc.  These parameters, combined with $\vcirc = 220$ km s$^{-1}$, predict an escape velocity of 584 km s$^{-1}$, which falls near the middle of the ranges proposed in Refs.~\cite{LT90, Rave06}.

The isotropic Maxwellian velocity distribution given by Eq.~(\ref{maxwell}) specifies the dark matter particles' motion relative to the halo.  However, we are interested in their motion relative to the XQC detector: $\vec{v}_\mathrm{observed} = \vec{v}_\mathrm{dm} - \vec{v}_\mathrm{detector}$ where the latter two velocities are measured with respect to the halo.  The velocity of the detector with respect to the halo has three components: the velocity of the Sun relative to halo, the velocity of the Earth with respect to the Sun, and the velocity of the detector with respect to the Earth.  

When discussing these velocities, it is useful to define a Galactic Cartesian coordinate system.  In Galactic coordinates, the Sun is located at the origin, and the $xy$-plane is defined by the Galactic disk.  The $x$-axis points toward the center of the Galaxy, and the $y$-axis points in the direction of the Sun's tangential velocity as it revolves around the Galactic center.  The $z$-axis points toward the north Galactic pole and is antiparallel to the angular momentum of the rotating disk. The motion of the Sun through the halo has two components.  First, there is the Sun's rotational velocity as it orbits the Galactic center: $\vcirc$ in the $y$ direction.  Second, there is the motion of the Sun relative to the spiral disk \cite{DB98}: $\vec{v}_\mathrm{\odot} = (10.00 \pm 0.36, 5.25\pm 0.62, 7.17 \pm 0.38)$ km s$^{-1}$ in Galactic Cartesian coordinates.  When the Earth's motion through the Solar System during its annual orbit of the Sun is expressed in Galactic coordinates \cite{Lewin}, the resulting velocity at the time of the XQC experiment (7.3 days after the vernal equinox) is $\vec{v}_\mathrm{Earth} = (29.14, 5.330, -3.597)$ km s$^{-1}$.

The final consideration is the velocity of the detector relative to the Earth.  The maximum velocity attained by the XQC rocket was less than 1.2 km s$^{-1}$.  This velocity is insignificant compared to the motion of the Sun relative to the halo. Moreover, the XQC detector collected data while the rocket rose and while it fell, and the average velocity of the rocket was only $0.104$ km s$^{-1}$.  Therefore, we neglect the motion of the rocket in the calculation of the dark matter wind.  Combining the motion of the Sun and the Earth then gives the total velocity of the XQC detector with respect to the halo during the experiment in Galactic Cartesian coordinates: $\vec{v}_\mathrm{detector} = (39.14 \pm 0.36, 230.5 \pm 30, 3.573 \pm 0.38)$ km s$^{-1}$.  Subtracting the velocity vector of the detector relative to the halo from the velocity vector of the dark matter relative to the halo gives the dark matter's velocity relative to the detector in Galactic coordinates.  However, we want the dark matter particles' velocities in the coordinate frame defined by the detector, where the $z$-axis is the field-of-view vector.  The XQC field of view was centered on $l = 90^\circ$, $b = +60^\circ$ in Galactic latitude and longitude \cite{big}, so the rotation from Galactic coordinates to detector coordinates may be described as a clockwise $30^\circ$ rotation of the $z$-axis around the $x$-axis, which is taken to be the same in both coordinate systems. 

\subsection{Dark Matter Interactions}
\label{sec:darkmatterinteractions}
Calorimetry measures the kinetic energy transferred from the dark matter to the absorbing material without regard for the specific mechanism of the scattering or any other interactions.  Consequently, the dark matter detection rate for a calorimeter depends only on the mass of the dark matter particle and the total cross section for elastic scattering between the dark matter particle and an atomic nucleus of mass number $A$, which is proportional to the cross section for dark matter interactions with a single nucleon ($\sn$).  The calorimeter measures the recoil energy of the target nucleus (mass $\mnuc$),
\begin{equation}
E_\mathrm{rec} = \left(\frac{1}{2}\mdm v_\mathrm{dm}^2\right)\frac{2\mnuc\mdm}{(\mnuc+\mdm)^2}(1-\cos \theta_\mathrm{CM}),
\label{Erec}
\end{equation}
where $\mdm$ and $v_\mathrm{dm}$ are dark matter particle's mass and velocity prior to the collision in rest frame of the target nucleus and $\theta_\mathrm{CM}$ is the scattering angle in the center-of-mass frame.  

If the momentum transferred to the nucleus, $q^2=2\mnuc E_\mathrm{rec}$, is small enough that the corresponding de Broglie wavelength is larger than the radius $R$ of the nucleus ($qR\ll \hbar$), then the scattering is coherent. In coherent scattering, the scattering amplitudes for each individual component in the conglomerate body are added prior to the calculation of the cross section, so the total cross section is proportional to the square of the mass number of the target nucleus.  Including kinematic factors \cite{GW85, Stark}, the cross section for coherent scattering off a nucleus is given by
\begin{equation}
\scoh(A) = A^2\left[\frac{\mred(\mathrm{DM,Nuc})}{\mred(\mathrm{DM,n})}\right]^2 \sn,
\label{coherentsig}
\end{equation}
where $\mred(\mathrm{DM,Nuc})$ is the reduced mass of the nucleus and the dark matter particle, $\mred(\mathrm{DM,n})$ is the reduced mass of a nucleon and the dark matter particle, and $A$ is the mass number of the nucleus.  Coherent scattering is isotropic in the center-of-mass frame of the collision.

Dark matter particles may be massive and fast-moving enough that the scattering is not completely coherent when the target nucleus is large \cite{Gould87}.  When the scattering is incoherent, the dark matter particle ``sees'' the internal structure of the nucleus, and the cross section for scattering is reduced by a ``form factor,'' which is a function of the momentum transferred to the nucleus during the collision ($q$) and the nuclear radius ($R$):
\begin{equation}
\frac{\deriv \sigma}{\deriv \Omega} = \frac{\scoh}{4\pi} F^2(q,R).
\label{defformfactor}
\end{equation} 
Since $q$ depends on the recoil energy, which in turn depends on the scattering angle, incoherent scattering is not isotropic.  

In this discussion of coherence, we have neglected the possible effects of the dark matter particle's internal structure by assuming that $\sn$ is independent of recoil energy.  If the dark matter particle is not point-like then $\sn$ decreases as the recoil momentum increases due to a loss of coherence within the dark matter particle.  Incoherence within the dark matter particle has observational consequences \cite{GKN02}, but these effects depend on the size of the dark matter particle.  To avoid restricting ourselves to a particular dark matter model, we assume that the dark matter particle is small enough that nucleon scattering is always coherent; when we discuss incoherence, we are referring to the effects of the nucleus's internal structure. 

According to the Born approximation, the form factor for nuclear scattering defined in Eq.~(\ref{defformfactor}) is the Fourier transform of the nuclear ground-state mass density \cite{Engel91, Lewin}.  The most common choice for the form factor \cite{Freedman74, Gould87} is $F^2(q,R) = \exp[-(qR_\mathrm{rms})^2/(3\hbar^2)]$, where $R_\mathrm{rms}$ is the root-mean-square radius of the nucleus.  For a solid sphere, $R_\mathrm{rms}^2 = (3/5) R^2$, so this form factor is equivalent to the form factor used in Ref.~\cite{ZF05}.  This form factor is an accurate approximation of the Fourier transform of a solid sphere for $(qR)/\hbar\lsim2$, but it grossly underestimates the reduction in $\sigma$ for larger values of $q$ \cite{Lewin}.  The maximum speed of a dark matter particle with respect to the XQC detector is $\sim800$ km s$^{-1}$ (escape velocity + detector velocity), and at that speed, the maximum possible value of $qR/\hbar$ for a collision with a Hg nucleus ($A=200$) is nearly ten for a 100 GeV dark matter particle, and the maximum possible value of $qR/\hbar$ increases as the mass of the dark matter particle increases.  Clearly, this approximation is not appropriate for a large portion of the dark matter parameter space probed by the XQC experiment.  

Furthermore, a solid sphere is not a very realistic model of the nucleus.  A more accurate model of the nuclear mass density is $\rho({\bf{r}}) = \int \mathrm{d}^3 r^\prime \rho_0({\bf{r^\prime}})\rho_1({\bf{r-r^\prime}})$, where $\rho_0$ is constant inside a radius $R_0^2 = R^2 - 5s^2$ and zero beyond that radius and $\rho_1 = \exp[-r^2/(2s^2)]$, where $s$ is a ``skin thickness" for the nucleus \cite{Helm56}.  The resulting form factor is
\begin{eqnarray}
F(q,R) = &&3 \left[\frac{\sin(qR/\hbar)-(qR/\hbar)\cos(qR/\hbar)}{(qR/\hbar)^3}\right]\nonumber\\
&&\times \mbox{  }\exp\left[-\frac{1}{2}(qs/\hbar)^2\right].
\label{exactF}
\end{eqnarray}
We follow Ref.~\cite{Lewin} in setting the parameters in Eq.~(\ref{exactF}): $s = 0.9$ fm and
\begin{equation}
R^2 = [(1.23 A^{1/3} - 0.6)^2 + 0.631 \pi^2 - 5s^2 ]\mbox{ fm$^2$},
\label{Rnuc}
\end{equation}
where $A$ is the mass number of the target nucleus. 

Despite its simple analytic form, the form factor given by Eq.~(\ref{exactF}) is computationally costly to evaluate repeatedly.  We use an approximation:
\begin{equation}
F^2 = \left\{ \begin{array}{ll} 
                        \exp\left[{-\left(\frac{qR}{\hbar}\right)^2\left\{{\frac{1}{5}
			  +\left(\frac{0.9 \mbox{ fm}}{R}\right)^2}\right\}}\right]& 
			\mbox{ if $\frac{qR}{\hbar} \leq 2$},\\
			 \frac{9(0.81)}{(qR/\hbar)^4}\exp\left[{-\left(\frac{qR}{\hbar}\right)^2
			     \left(\frac{0.9 \mbox{ fm}}{R}\right)^2}\right]& 
			 \mbox{ if  $\frac{qR}{\hbar} > 2$}.
                        \end{array}  \right.
\label{approxF}
\end{equation}
The low-$q$ approximation combines the standard approximation for the solid sphere with the factor accounting for the skin depth of the nucleus.  The high-$q$ approximation was derived from the asymptotic form of the first spherical Bessel function and normalized so that the total cross section is as close as possible to the exact result.  The error in the total cross section due to the use of the approximation is less than 1\% for nearly all dark matter masses; the sole exception is $\mdm \sim 10-100$ GeV, and even then the error is less than 5\%.  Unless otherwise noted, we use this approximation for the form factor throughout this analysis.  We also assume that the dark matter particle does not interact with nuclei in any way other than elastic scattering.
\section{Analysis of XQC Constraints}
\label{sec:analysis}
To obtain an accurate description of the XQC experiment's ability to detect strongly interacting dark matter particles, we turned to Monte Carlo simulations.  The Monte Carlo code we wrote to analyze the XQC experiment simulates a dark matter particle's journey through the atmosphere to the XQC detector chamber, its path through the detector chamber to a calorimeter, and its interaction with the sensitive components of the calorimeter.  This latter portion of the code also records how much energy the particle deposits in the calorimeter through scattering.  The results of several such simulations for the same set of dark matter properties may be used to predict the likelihood that a given dark matter particle will deposit a particular amount of energy into the calorimeter.  These probabilities of various energy deposits predict the recoil-energy spectrum the XQC detector would observe if the dark matter particles have a given mass and nucleon-scattering cross section.  This simulated spectrum may then be compared to the XQC data to find which dark matter parameters are excluded by the XQC experiment.

\subsection{Generating Simulated Energy-Recoil Spectra}
\label{sec:MonteCarlo}
The basic subroutine in our Monte Carlo algorithm is the step procedure.  The step procedure begins with a particle with a certain velocity vector and position in a given material and moves the particle a certain distance in the material, returning its new position and velocity.  The step procedure also determines whether or not a scattering event occurred during the particle's trek and updates the velocity accordingly.  The number of expected collisions in a step of length $l$ through a material with target number density $n$ is $n\times\stot\times l$, where $\stot$ is the total scattering cross section obtained by integrating Eq.~(\ref{defformfactor}) over the scattering angle, or equivalently, the recoil momentum $q$:
\begin{equation}
\stot = \frac{\scoh}{q_\mathrm{max}^2}\int_0^{q_\mathrm{max}^2} F^2(q,R) \intderiv q^2,
\label{sigtot}
\end{equation}
where $q_\mathrm{max}$ is the maximum possible recoil momentum.  The step length $l$ is chosen so that it is at most a tenth of the mean free path through the material, so the number of expected collisions is less than one and represents the probability of a collision.  After each step, a random number between zero and one is generated using the ``Mersenne Twister'' (MT) algorithm \cite{rng} and if that random number is less than the probability of a collision, the particle's energy and trajectory are updated.  First, a recoil momentum is selected according to the probability distribution $P(q^2) = F^2(q,R) \scoh/(q_\mathrm{max}^2\stot)$, where the exact form factor is used for $qR/\hbar >2$ so that the oscillatory nature of the form factor is not lost.  The recoil momentum determines the recoil energy and the scattering angle in the center-of-mass frame through Eq.~(\ref{Erec}).  The scattering is axisymmetric around the scattering axis, so the azimuthal angle is assigned a random value between 0 and 2$\pi$.  The scattering angles are used to update the particle's trajectory, and its speed is decreased in accordance with the kinetic energy transferred to the target nucleus.  The step subroutine repeats until the particle exits the simulation, or its kinetic energy falls below 0.1 eV, or the energy deposited in the calorimeter exceeds the saturation point of 4000 eV. 

Our simulation treats the atmosphere as a 4.6$\times$4.6 cm square column with periodic boundary conditions, the bottom face of which covers the top of the conical detector chamber described in Section \ref{sec:XQC}. This implementation assumes that for every particle that exits one side of the column, there is a particle that enters the column from the opposite side with the same velocity.  The infinite extent of the atmosphere and its translational invariance makes this assumption reasonable.  The atmosphere column extends to an altitude 1000 km; increasing the atmosphere height beyond 1000 km has a negligible effect on the total number of collisions in the atmosphere.  The simulation begins with a dark matter particle at the top of the atmosphere column at a random initial position on the 4.6$\times$4.6 cm square.  Its initial velocity with respect to the dark matter halo is selected according to the isotropic Maxwellian velocity distribution function given by Eq.~(\ref{maxwell}), and then the velocity relative to the detector is found via the procedure described in Section \ref{sec:darkmatterwind}.  

The dark matter particle's path from the top of the atmosphere to the detector is modeled using the step procedure described above.  The simulation of the particle's interaction with the atmosphere ends if the particle's altitude exceeds 1000 km or if the particle falls below the height of the XQC rocket.  We use the time-averaged altitude ($201.747$ km) as the constant altitude of the rocket.  We made this simplification because it allows us to ignore the periodic inactivity of each calorimeter and treat the detector as thirty-four calorimeters that are active for 100.7 seconds.  When the dark matter particle hits the rocket, its path through the five filter layers is also modeled using the step procedure, as is its path through the calorimeters.  In addition to being smaller than the mean free path, the step length is chosen so that the particle's position relative to the boundaries of the detectors is accurately modeled.  The simulation ends when the dark matter particle's random-walk trajectory takes it out of the detector chamber.  As mentioned in Section \ref{sec:XQC}, the calorimeter detects the sum of all the recoil energies if the dark matter particle is scattered multiple times.

When the dark matter particle is unlikely to experience more than one collision in the calorimeter, this simulation is far more detailed than is required to accurately predict the energy deposited by the dark matter particle.  This is the case for the lightest ($\mdm \leq 10^2$ GeV) and weakest-interacting ($\sn \leq 10^{-26}$ cm$^2$) dark matter particles that the XQC calorimeters are capable of detecting.  Since the lightest dark matter particles are also the most numerous, many Monte Carlo trials are required to sample all the possible outcomes of a dark matter particle's encounter with the detector.  The simulation described above is too computationally intensive to run that many trials, so we used a faster and simpler simulation to model the interactions of these dark matter particles.  This simulation assumes that the particle will experience at most one collision in the atmosphere and at most one collision in each filter layer and each layer of the calorimeter. The simulation ends if the probability of two scattering events in either the atmosphere or any of the filter layers exceeds 0.1.  Instead of tracking the dark matter particle's path through the atmosphere, the total overburden for the atmosphere is used to determine the probability that the dark matter particle scatters in the atmosphere, and the particle only reaches the detector if its velocity vector points toward the detector after the one allowed scattering event.  Also, instead of the small step lengths required to accurately model the random walk of a strongly interacting particle, each layer is crossed with a single step.  These simplifications reduce the runtime of the simulation by a factor of 100, making it possible to run 10$^{10}$ trials in less than one day.

\subsection{Comparing the Simulations to the XQC Data}
\label{sec:comparison}

In order to compare the probability spectra produced by our Monte Carlo routine to the results of the XQC experiment, we must multiply the probabilities by the number of dark matter particles that are encountered by the initial surface of the Monte Carlo routine.  When the initial velocity of the dark matter particle is chosen, the initial velocity may point toward or away from the detector; in the latter case, the trial ends immediately.  Consequently, the Monte Carlo probability that the particle deposits no energy in the calorimeter already includes the probability that the dark matter particle does not have a halo trajectory that takes it into the atmosphere.  Therefore, the probabilities resulting from the Monte Carlo routine should be multiplied by the number of particles in the volume swept out by the initial 4.6$\times$4.6 cm$^2$ square surface during the 100.7$f(E)$ seconds of observation time, where $f$ is the fraction of the observing time that the XQC detector was sensitive to deposits of energy $E$.  For energies between 36 and 88 eV, $f$ is 0.5083, and the value of $f$ increases to one over energies between 88 and 128 eV.  The detector was also slightly sensitive to lower energies: between 29 and 35 eV, $f$ increases from 0.3815 to 0.5083.  The normal of the initial surface points along the detector's field of view, and the surface moves with the detector; using the detector velocity given in Section \ref{sec:darkmatterwind}, the number of dark matter particles encountered by the initial surface is $N_
\mathrm{dm}=f\times(\rho_\mathrm{dm}/\mdm)\times[(2.5 \pm 0.3)\times 10^{10}$ cm$^3]$, where $\rho_\mathrm{dm}$ is the local dark matter density.

\begin{figure}
\centerline{\epsfig{file=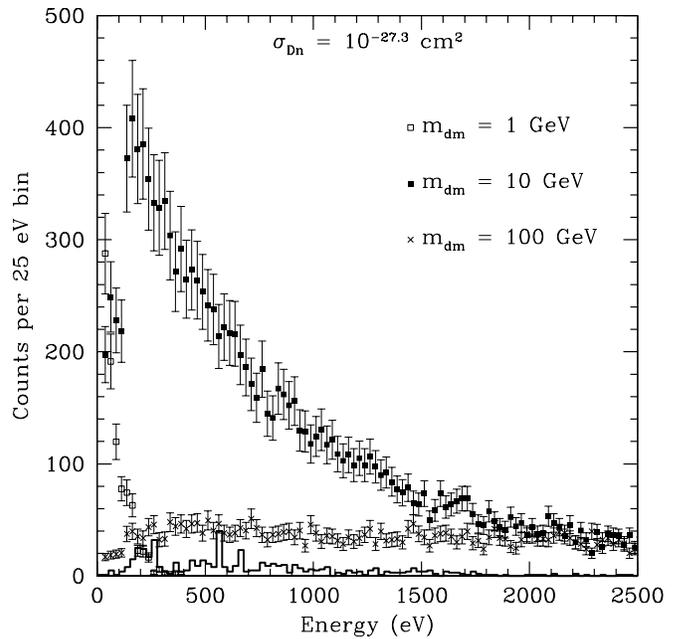, width=3.5in}}
\caption{Simulated event spectra for dark matter particles with masses of 1, 10 and 100 GeV and a total nucleon-scattering cross section of $10^{-27.3}$ cm$^2$.  In addition to the events depicted in these spectra, the simulations predict $1300 \pm 160$ events with energies greater than 4000 eV when $\mdm=10$ GeV and $10,000 \pm 1200$ such events when $\mdm=100$ GeV.  The histogram represents the XQC observations.}
\label{fig:Sn3.3spectrum}
\end{figure}  

\begin{figure}
\centerline{\epsfig{file=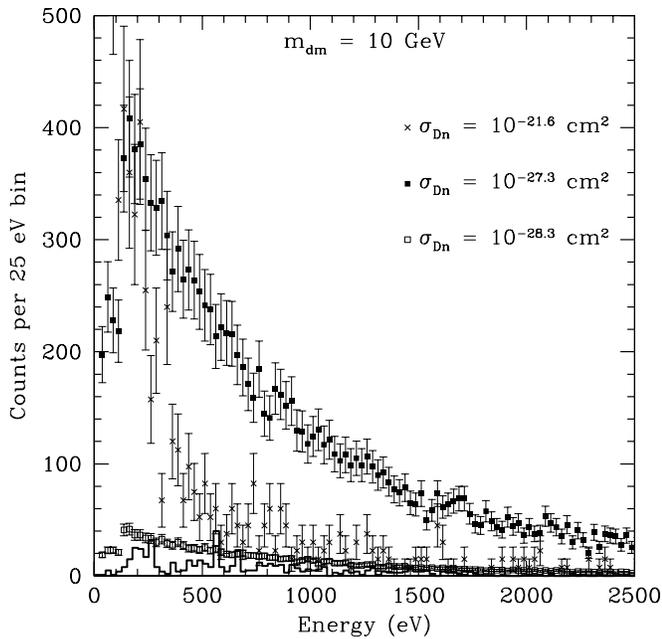, width=3.5in}}
\caption{Simulated event spectra for 10-GeV dark matter particles with total nucleon-scattering cross sections of $10^{-21.6}, 10^{-27.3}$ and $10^{-28.3}$ cm$^2$.  In addition to the events depicted in these spectra, the simulations predict $140 \pm 37$ events with energies greater than 4000 eV when $\sn = 10^{-21.6}$ cm$^2$, $1300 \pm 160$ such events when $\sn =10^{-27.3}$ cm$^2$, and $120 \pm 15$ such events when $\sn = 10^{-28.3}$ cm$^2$.  The histogram represents the XQC observations.}
\label{fig:M1spectrum}
\end{figure}  

The simulated event spectra produced by our Monte Carlo routine indicate that particles with masses less than 1 GeV very rarely deposit more than 100 eV inside the XQC calorimeters.   Conversely, particles with masses greater than 100 GeV nearly always deposit more than 4000 eV when they interact with the XQC calorimeters, so constraints on $\sn$ for these $\mdm$ values arise from the over-saturation ($E \geq 4000$ eV) event rate.  Fig.~\ref{fig:Sn3.3spectrum} shows simulated spectra for three $\mdm$ values that lie between these two extremes, along with a histogram that depicts the XQC observations.  Given an initial velocity of 300 km s$^{-1}$ relative to the XQC detector, a 1-GeV particle can only deposit up to 66 eV in a single collision with an Si nucleus, so the spectrum for these particles is confined to very low energies.   Meanwhile, a 10-GeV particle and a 100-GeV particle with the same initial velocity can deposit up to 900 eV and 44,000 eV, respectively, in a single collision with an Hg nucleus.  In fact, ignoring any loss of coherence, all recoil energies between 0 and 44,000 eV are equally likely during a collision between an Hg nucleus and a 100-GeV dark matter particle.  That's why the $\mdm=100$ GeV spectrum in Fig.~\ref{fig:Sn3.3spectrum} is flat below 2500 eV and why the simulations predict 10,000 events with energies greater than 4000 eV for this value of $\mdm$ and $\sn$. 

Fig.~\ref{fig:M1spectrum} shows how changing the total cross section for elastic scattering off a nucleon affects the simulated spectra generated by our Monte Carlo routine for a single dark matter particle mass ($\mdm=10$ GeV).  We see that increasing $\sn$ from 10$^{-28.3}$ cm$^2$ to 10$^{-27.3}$ cm$^2$ increases all of the counts by a factor of ten but leaves the basic shape of the spectrum unchanged.   For much larger values of $\sn$, however, the particle loses a considerable amount of its energy while traveling through the atmosphere.  Consequently, high-energy recoil events become less frequent, as shown by the spectrum for $\sn = 10^{-21.6}$ cm$^2$.  For larger values of $\sn$, too much energy is lost in the atmosphere for the particle to be detectable by the XQC experiment.
\begin{table}
\begin{center}
\begin{ruledtabular}
\begin{tabular}{|lc|lc|}
\hline 
Energy Range (eV) & Counts & Energy Range (eV)& Counts \\
\hline
29 - 36 & 0 & 945 - 1100 & 31 \\
36 - 128 & 11 & 1100 - 1310 & 30 \\
128 - 300 & 129 & 1310 - 1500 & 29 \\
300 - 540 & 80 & 1500 -1810 & 32 \\
540 - 700 & 90 & 1810 - 2505 & 15 \\
700 - 800 & 32 & $\geq$ 4000 & 60 \\
800 - 945 & 48 & & \\
\hline
\end{tabular}
\end{ruledtabular}
\caption{The binned XQC results used for comparison with our Monte Carlo simulations.}
\label{tab:bins}
\end{center}
\end{table}     

When comparing the simulated measurements to the XQC data, we group the events into the thirteen energy bins given in Table \ref{tab:bins}.  We generally use large bins because it reduces the fractional error in the probabilities generated by our Monte Carlo routine by increasing the probability of each bin: $\delta p_i/p_i = 1/\sqrt{p_i t}$, where $t$ is the number of trials and $p_i$ is the probability of an energy deposit in the $i$th bin.  Given that the number of trials is limited by runtime constraints, increasing the bin size is often the only way to obtain bin probabilities with $\delta p_i/p_i$ values much less than one.  When choosing our binning scheme, we attempted to maximize bin size while preserving as many features of the observed spectrum as possible.  We also grouped all energies for which $f\neq1$ into  two bins; we ignore the variation in $f$ within these bins and set $f = 0.3815$ in the lowest-energy bin and $f = 0.5083$ in the next-to-lowest bin.

Unfortunately, we do not know the number of X-ray events in any of the bins listed in Table \ref{tab:bins}.  We considered using a model to subtract off the X-ray background but, given any model's questionable accuracy, we decided not to use it in our analysis.  Our ignorance of the X-ray background forces us to treat the number of observed counts in each bin as an upper limit on the number of dark matter events in that energy range.  Consequently, we define a parameter $X^2$ that measures the extent of the discrepancy between the simulated results for a given $\mdm$ and $\sn$ and the XQC observations while ignoring bins in which the observed event count exceeds the predicted contribution from dark matter:
\begin{equation}
X^2 \equiv \sum_{i=1}^{i=\mbox{\# of Bins}} \left\{\frac{(E_i-U_i)^2}{E_i} \mbox{ with } U_i<E_i \right\},
\label{xdef}
\end{equation}
where $E_i=N_\mathrm{dm}\times p_i$ is the number of counts in the $i$th bin predicted by the Monte Carlo simulation and $U_i$ is the number of observed counts in the same bin.  We use a second Monte Carlo routine to determine how likely it is that a set of observations would give a value of $X^2$ as large or larger than the one derived from the XQC data given a mean signal described by the set of $E_i$ derived from the simulation Monte Carlo.  

In the comparison Monte Carlo, a trial begins by generating a new set of $E_i$ by sampling the error distributions of $N_\mathrm{dm}$ and $p_i$.  The distribution of $N_\mathrm{dm}$ values is assumed to be Gaussian with the mean and standard deviation given above.  The probability $p_i$ is derived from $p_i \times t$ events in the simulation Monte Carlo (recall that $t$ is the number of trials), so a new value for $p_i$ is generated by sampling a Poisson distribution with a mean of $p_i \times t$ and dividing the resulting number by $t$.  Once a new set of $E_i$ has been found, the routine generates a simulated number of observed counts for each bin according to a Poisson distribution with a mean of $E_i$.  The value of the $X^2$ parameter for the new $E_i$ and $U_i$ is computed and compared to the value for the original $E_i$ and the XQC observations, ${X^2}_\mathrm{XQC}$.  The number of trials needed to accurately measure the probability P(X) that $X^2\geq{X^2}_\mathrm{XQC}$ is determined by requiring that the variation in the mean value of $X^2$ over ten Monte Carlo simulations does not exceed (100-$C$)\%, where $C$\% is the desired confidence level and that the range P(X)$\pm$(5$\times$ the variation in P(X)) does not contain $(100-C)/100$.

\section{Results and Discussion}
\label{sec:results}

\begin{figure}
\centerline{\epsfig{file=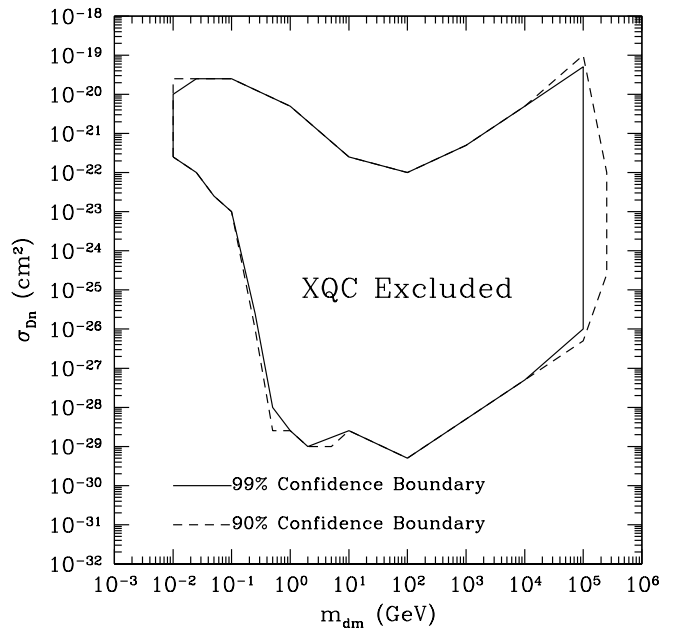, width=3.55in}}
\caption{The region of dark matter parameter space excluded by the XQC experiment; $\sn$ is the total cross section for scattering off a nucleon and $\mdm$ is the mass of the dark matter particle.  This exclusion region follows from the assumption that the local dark matter density is 0.3 GeV cm$^{-3}$ and that all of the dark matter shares the same value of $\sn$.}
\label{fig:result}
\end{figure}    

The XQC experiment rules out the enclosed region in $(\mdm, \sn)$ parameter space shown in Fig.~\ref{fig:result}.  The overburden from the atmosphere and the filtering layers assures that there will be a limit to how strongly a dark matter particle can interact with baryons and still reach the XQC calorimeters; this overburden is responsible for the top edge of the exclusion region.  Conversely, if $\sn$ is too small, the dark matter particles will pass through the calorimeters without interacting.  The low-energy threshold of the XQC calorimeters places a lower bound on the excluded dark matter particle masses; if $\mdm$ is too small, then the recoil energies are undetectable.  On the other side of the mass range, the XQC detector is not sensitive to $\mdm \gsim 10^5$ GeV because the number density of such massive dark matter particles is too small for the XQC experiment to detect.

The exclusion region shown in Fig.~\ref{fig:result} has a complicated shape, but its features are readily explicable.  As $\mdm$ increases, the range of excluded $\sn$ values shifts to lower values and then moves up again.  The downward shift for $\mdm$ between 0.1 GeV and 100 GeV is due to the effects of coherent nuclear scattering. Since $\scoh$ increases with increasing $\mdm$ for fixed $\sn$, a 100-GeV particle interacts more strongly in the atmosphere and in the detector than a 1-GeV particle with the same $\sn$.  Consequently, both the upper and lower boundaries of the excluded region decrease with increasing mass for $\mdm\lsim100$ GeV.  The scattering of dark matter particles with larger masses is incoherent, and the form factor discussed in Section \ref{sec:darkmatterinteractions} causes $\stot$ to decrease as mass increases for fixed $\sn$.  Moreover, particles that are more massive than the target nuclei have straighter trajectories than lighter dark matter particles due to smaller scattering angles in the detector rest frame.  The loss of coherence also contributes because incoherent scattering makes small scattering angles more probable.  A straight trajectory is shorter than a random walk, so the more massive particles interact less in the atmosphere and the detector than the more easily-deflected lighter particles.  Due to both of these effects, the upper and lower boundaries of the exclusion region increase with increasing $\mdm$ for $\mdm\gsim100$ GeV.

The lower left corner of the exclusion region also has two interesting features.  First, the lower bound on the excluded value of $\sn$ decreases sharply as $\mdm$ increases from 0.1 GeV to 0.5 GeV.  A dark matter particle with the maximum possible velocity with respect to the detector (800 km s$^{-1}$) must have a mass greater than $0.24$ GeV to be capable of depositing 29 eV in the calorimeter in a single collision.  Lighter particles are only detectable if they scatter multiple times inside the calorimeter, and multiple scatters require a higher value of $\sn$.  Since their analysis does not allow multiple collisions, the XQC exclusion region found in Ref.~\cite{ZF05} does not extend to masses lower than 0.3 GeV for any value of $\sn$.  Second, there is a kink in the lower boundary at $\mdm = 10$ GeV; the constraint on $\sn$ is not as strong for this mass.  The simulated spectra produced by our Monte Carlo routine for $\mdm = 10$ GeV and $\sn \lsim 10^{-25}$ cm$^2$ reveal that the particle is most likely to deposit between 100 and 600 eV, as exemplified by the spectra depicted in Fig.~\ref{fig:M1spectrum}.  The background in this energy range is very high, so the XQC constraints are not as strict at these energies.

\begin{figure}
\centerline{\epsfig{file=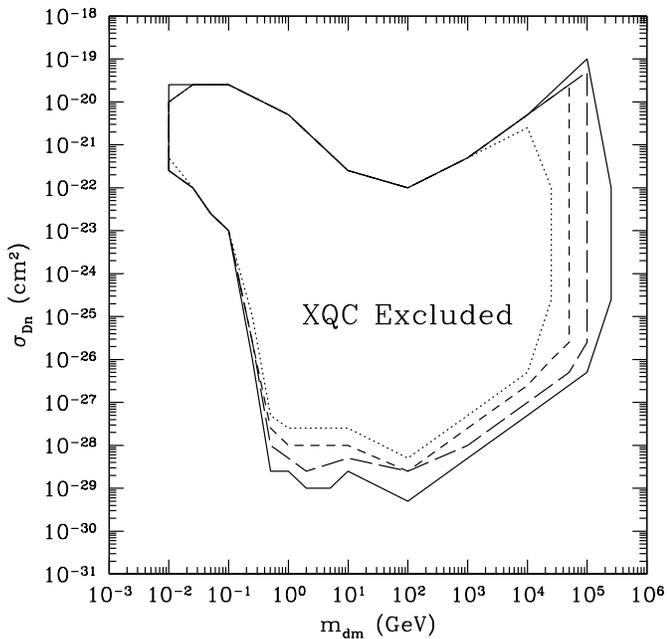, width=3.55in}}
\caption{The region of dark matter parameter space excluded to 90\% confidence by the XQC experiment for several values of the local density of dark matter with a total nucleon scattering cross section $\sn$ and mass $\mdm$.  The four densities shown are 0.3 GeV cm$^{-3}$ (solid line), 0.15 GeV cm$^{-3}$ (long dashed line), 0.075 GeV cm$^{-3}$ (short dashed line) and 0.03 GeV cm$^{-3}$ (dotted line).}
\label{fig:densities}
\end{figure}  

Altering the local density of dark matter that strongly interacts with baryons changes the exclusion region.  Fig.~\ref{fig:densities} shows the 90\%-confidence exclusion regions for four values of the local density of dark matter particles with strong baryon interactions: 0.3 GeV cm$^{-3}$ (solid line), 0.15 GeV cm$^{-3}$ (long dashed line), 0.075 GeV cm$^{-3}$ (short dashed line) and 0.03 GeV cm$^{-3}$ (dotted line).  These different local densities could arise due to variations in the local dark matter density due to mini-halos or streams.  They also describe models where the dark matter does not consist of a single particle species and the dark matter that strongly interacts with baryons is a fraction $f_d$ of the local dark matter.  In that case, the four exclusion regions in Fig.~\ref{fig:densities} correspond to $f_d$ =1, 0.5, 0.25, and 0.1.   

\begin{figure*}
\includegraphics[width=5.5in]{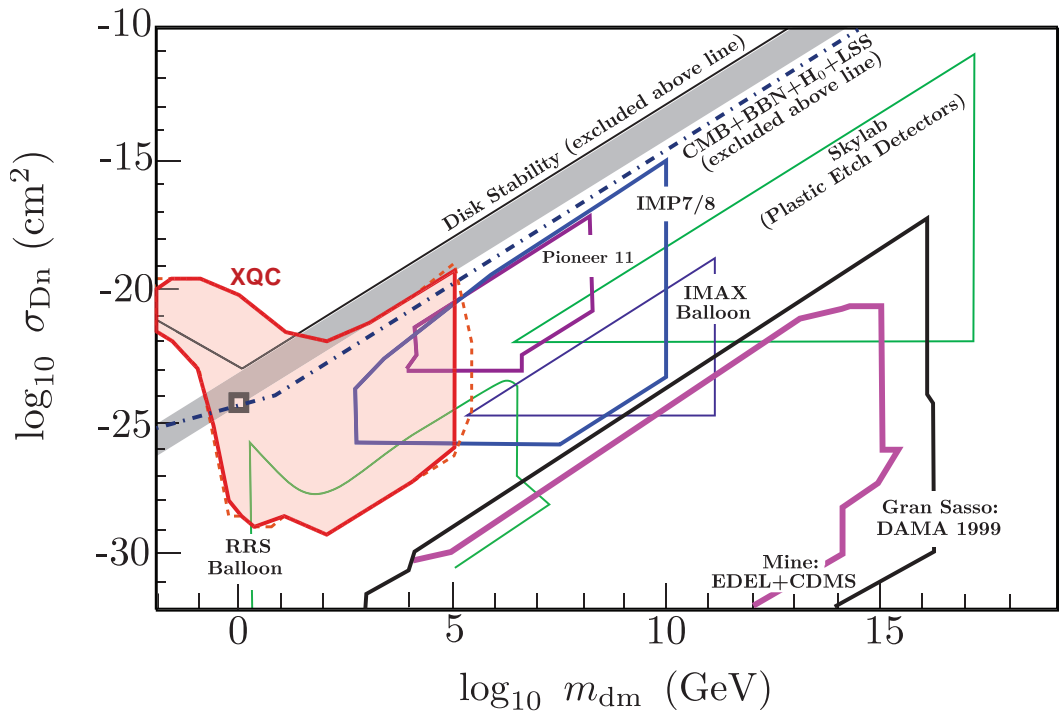}
\caption{Plot of the scattering cross section for dark matter particles and nucleons ($\sn$) versus dark matter particle mass ($\mdm$) showing the new XQC limits along with other current experimental limits.  The red XQC exclusion region is the same as shown in Fig.~\ref{fig:result}, and the other experiments are discussed in the text.  The dark gray region shows the maximal range of dark matter self-interaction cross section consistent with the strongly self-interacting dark matter model of structure formation \cite{oriSIDM, Wandelt}.  The square marks the value of the scattering cross section for neutron-nucleon interactions.}
\label{fig:allbounds}
\end{figure*}

Fig.~\ref{fig:densities} indicates that the top and left boundaries of the XQC exclusion region are not highly sensitive to the dark matter density.  In particular, the upper left corner of the exclusion region ($0.01 \leq \mdm \leq 0.1$ GeV) is nearly unaffected by lowering the dark matter density.  This consistency indicates our Monte Carlo-generated exclusion region is smaller than the true exclusion region in this corner.  If the dark matter is light ($\mdm \lsim 0.1$ GeV), then the number of dark matter particles encountered by the XQC detector is very large ($N_\mathrm{dm} \gsim 7\times10^{10}$).  As previously mentioned, the upper left corner of the XQC exclusion region results from multiple scattering events, so the simpler version of our Monte Carlo code described in Section \ref{sec:MonteCarlo} is not applicable.  Consequently, it is not possible to run more than $10^9$ trials in a week, so each scattering event in the simulation corresponds to more than one scattering event in the detector for all the densities shown in Fig.~\ref{fig:densities}.  Therefore, decreasing the density does not change the result.  If it were possible to run $10^{11}$ trials, then the upper left corner of the exclusion region would expand and differences between the different density contours would emerge.   Since the upper left corner of the XQC exclusion region is already ruled out by astrophysical constraints (see Fig.~\ref{fig:allbounds}), we have not invested in the computational time necessary to expand this corner.

The upper boundary of the exclusion region is also not greatly affected by decreasing the particle density, even when $N_\mathrm{dm} $ is small enough that the Monte Carlo routine is capable of running more than $N_\mathrm{dm} $ trials ($\mdm \geq 100$ GeV).  This robustness indicates that the overburden of the XQC experiment effectively prevents all dark matter particles with $\sn$ values greater than the upper bound of the exclusion region from reaching the detector, so it does not matter how many particles are encountered.  Finally, as discussed previously, the lower portion of the exclusion region's left boundary ($\sn \leq 10^{-23}$ cm$^2$) is set by the energy threshold for detection and is therefore independent of $N_\mathrm{dm} $.

Examining the features of the excluded region allows us to predict how the region may be expanded by a future XQC-like experiment.  Decreasing the overburden by either increasing the rocket's altitude or reducing the filtering will push the top boundary of the excluded region upwards.  Decreasing the energy detection threshold will extend the excluded region to lower masses.  It may also extend the exclusion region to higher values of $\sn$ for all masses since strongly interacting particles lose much of their energy in the atmosphere and arrive at the calorimeter with too little energy to produce a detectable signal.  Increasing the size or number of calorimeters would increase the sensitivity and extend the excluded region to lower values of $\sn$.  Finally, increasing the observation time would increase $N_\mathrm{dm}$, and that would extend the right and bottom boundaries of the excluded region.  

\section{Conclusion}
\label{sec:conclusion}

The X-ray Quantum Calorimetry (XQC) experiment is a powerful detector of dark matter that interacts strongly with baryons due to its high altitude and minimal shielding.  The XQC measurements rule out a large range of hitherto unconstrained dark matter masses and scattering cross sections.Ê The excluded range was first derived in Refs.~\cite{Wandelt, MS01} based on rough analytic estimates. In this paper, we have improved upon these results using detailed Monte Carlo simulations to predict how a dark matter particle of a given mass and cross section for nucleon scattering would interact with the XQC calorimeters.  Unlike Ref.~\cite{ZF05}, our analysis includes the atmosphere and the shielding of the detector, so our result includes the upper limit on excluded $\sn$ values, which had not yet been accurately determined.  Our simulation also models the internal geometry of the XQC detector and the random walk of particles through it, which is not possible using the analytical approaches of Refs.~\cite{Wandelt, MS01, ZF05}.  

The resulting exclusion region is significantly different than its analytical predecessors.  When multiple scatterings are included, the XQC experiment is sensitive to dark matter particles with masses below 0.3 GeV and cross sections for nucleon scattering between 10$^{-24}$ and 10$^{-20}$ cm$^2$.  Unlike Ref.~\cite{ZF05}, we find that the XQC exclusion region does not include $\sn < 10^{-29}$ cm$^2$ for dark matter masses less than 10 GeV.  Ref.~\cite{ZF05} obtained a more restrictive upper bound because they assumed a specific X-ray background while we treat all events as potential dark matter interactions.  At higher masses, the lower boundary of our exclusion region is much higher than in Refs.~\cite{Wandelt, MS01} because they over-estimated the XQC sensitivity by assuming coherent scattering.  It also appears that Refs.~\cite{Wandelt, MS01} underestimated the atmospheric and shielding overburden for the XQC detector because our exclusion region does not extend to values of $\sn$ as large as those included in their exclusion region.  We also assume a lower local dark matter density than Refs.~\cite{Wandelt, MS01} (0.3 instead of 0.4 GeV cm$^{-3}$), so some of the shrinkage of the exclusion region may be attributed to the reduction in the assumed number density of dark matter particles.

Fig.~\ref{fig:allbounds} shows how the XQC exclusion region depicted in Fig.~\ref{fig:result} complements the exclusion regions from other experiments that are sensitive to similar values of $\sn$ and $\mdm$.  For a summary of some of the other experimental constraints as of 1994, see Ref.~\cite{McGuire94}.  The constraints to $\sn$ from Pioneer 11 \cite{Simpson80}, Skylab \cite{SP78}, and IMP7/8 \cite{IMP01} were interpreted by Refs.~\cite{McGuire94, Wandelt, MS01}.   There have been two balloon-borne searches for dark matter, the IMAX experiment  \cite{McGuire94, McGuire95} and the Rich, Rocchia \& Spiro (RRS) \cite{RRS87} experiment.  Although underground detectors are designed to detect WIMPs, DAMA \cite{Bacci94, Bernabei99} does exclude $\sn$ values within the range of interest, and relevant constraints may be derived from Edelweiss (EDEL) and CDMS \cite{AB93, AB93E}.  

All of the exclusion regions shown in Fig.~\ref{fig:allbounds} were derived assuming that all the dark matter is strongly interacting.  A local dark matter density of 0.4 GeV cm$^{-3}$ was assumed in the analysis of the exclusion regions from Pioneer 11, Skylab and the RRS experiment, while all the other exclusion regions were derived assuming a local dark matter density of 0.3  GeV cm$^{-3}$.  Furthermore, the derivations of all the shown exclusion regions other than the XQC region and the EDEL+CDMS region assume that the scattering between dark matter particles and nuclei is coherent.  Therefore, these exclusion regions are likely too broad because they over-estimate the cross section for nuclear scattering.  A comparison of the XQC exclusion region reported in Refs.~\cite{Wandelt, MS01} and our exclusion region indicates that assuming coherent scattering extends the exclusion region for $\mdm \geq 1000$ GeV to $\sn$ values that are roughly $A\times$ smaller than the lower boundary of our exclusion region, where $A$ is the mass number of the largest target nucleus.

Fig.~\ref{fig:allbounds} also shows the bound on $\sn$ from the CMB and large-scale structure (LSS) obtained when one assumes prior knowledge of the Hubble constant $H_0$ and the cosmic baryon fraction (from BBN) \cite{CHS02}.  This bound is nominally stronger than the bound from disk stability \cite{Stark}, but it is less direct in that it requires combining different measurements and depends on the cosmological model; consequently, we show both bounds in Fig.~\ref{fig:allbounds}.  Measurements of primordial element abundances give an upper limit of $\sn/\mdm \lsim 4 \times 10^{-16}$ cm$^2$ GeV$^{-1}$ \cite{CFPW02}.  Since this upper bound lies well beyond the upper bound from disk stability, we do not include it in Fig.~\ref{fig:allbounds}.  We also do not display the constraints from cosmic rays \cite{CFPW02} because they are derived from inelastic interactions that are model-dependent.  

As shown in Fig.~\ref{fig:allbounds}, the XQC experiment rules out a wide region of ($\mdm, \sn$) parameter space that was not probed by prior dark matter searches.  Of particular interest is the darkly shaded range of $\sn$ values that corresponds to the maximal range of dark matter self-interaction cross sections consistent with the strongly self-interacting dark matter model of structure formation \cite{oriSIDM, Wandelt}.  If the dark matter consists of exotic hadrons whose interactions with nucleons are comparable to their self-interactions, then $\sn$ for these particles would lie in or near the darkly shaded region in Fig.~\ref{fig:allbounds}.  Previous estimates of the XQC exclusion region \cite{Wandelt, MS01} indicated that the XQC experiment rules out all the darkly shaded $\sn$ values for $1 \lsim \mdm \lsim 10^4$ GeV.  Our analysis reveals that this is not the case; portions of the darkly shaded region for $\mdm \gsim 20$ GeV are not excluded by the XQC experiment, although they are ruled out by observations of LSS and the CMB.  The mass-$\sigma$ combination corresponding to nucleon-neutron scattering (the square in Fig.~\ref{fig:allbounds}) lies within the exclusion region of the XQC experiment, and the only portion of the darkly shaded region that is unconstrained corresponds to dark matter masses smaller than 0.25 GeV.  

It is important to note, however, that the cross section for dark matter self-interactions need not be comparable to the cross section for nucleon scattering;  $\sn$ could differ by a few orders of magnitude from the self-interaction cross section (as is the case for Q-balls).   Furthermore, no interactions with baryons are required for self-interacting dark matter to resolve the tension between the collisionless dark matter model and observations of small-scale structure.

Another XQC detector is scheduled to launch in the upcoming year.  This experiment will have twice the observing time of the XQC experiment used in this analysis.  As discussed in Section \ref{sec:results}, increasing the observing time will extend the exclusion region to higher masses and weaker interactions.   The future XQC experiment will also have a lower energy threshold (15 eV) and will maintain sensitivity to all energies above this threshold throughout the run.  The increased sensitivity to low energies will shift the lower ($\sn \leq 10^{-23}$ cm$^2$) left boundary of the exclusion region to lower masses.  A lower energy threshold of 15 eV will make the experiment sensitive to single recoil events involving dark matter particles more massive than 0.17 GeV, as discussed in Section \ref{sec:results}.   Clearly, the next-generation XQC experiment will be an even more powerful probe of interactions between dark matter particles and baryons than its predecessor.  

\begin{acknowledgments}
A. L. E. would like to thank Robert Lupton and Michael Ramsey-Musolf for useful discussions.  D. M. thanks the Wallops Flight Facility launch support team and the many undergraduate and graduate students that made this pioneering experiment possible.  The authors also thank Randy Gladstone for his assistance with the atmosphere model.   A. L. E. acknowledges the support of an NSF Graduate Fellowship.  P. J. S. is supported in part by US Department of Energy grant DE-FG02-91ER40671.  P. C. M. acknowledges current support for this project from a Robert M. Walker Senior Research Fellowship in Experimental Space Science from the McDonnell Center for the Space Sciences, as well as prior institutional support for this project from the Instituto Nacional de T{\'e}cnica Aeroespacial (INTA) in Spain, from the University of Bielefeld in Germany, and from the University of Arizona.
\end{acknowledgments}




\begin{thebibliography}{87}
\expandafter\ifx\csname natexlab\endcsname\relax\def\natexlab#1{#1}\fi
\expandafter\ifx\csname bibnamefont\endcsname\relax
  \def\bibnamefont#1{#1}\fi
\expandafter\ifx\csname bibfnamefont\endcsname\relax
  \def\bibfnamefont#1{#1}\fi
\expandafter\ifx\csname citenamefont\endcsname\relax
  \def\citenamefont#1{#1}\fi
\expandafter\ifx\csname url\endcsname\relax
  \def\url#1{\texttt{#1}}\fi
\expandafter\ifx\csname urlprefix\endcsname\relax\def\urlprefix{URL }\fi
\providecommand{\bibinfo}[2]{#2}
\providecommand{\eprint}[2][]{\url{#2}}

\bibitem[{\citenamefont{{Rubin} et~al.}(1980)\citenamefont{{Rubin}, {Thonnard},
  and {Ford}}}]{Rot}
\bibinfo{author}{\bibfnamefont{V.~C.} \bibnamefont{{Rubin}}},
  \bibinfo{author}{\bibfnamefont{N.}~\bibnamefont{{Thonnard}}},
  \bibnamefont{and} \bibinfo{author}{\bibfnamefont{W.~K.}
  \bibnamefont{{Ford}}}, \bibinfo{journal}{The Astrophysical Journal}
  \textbf{\bibinfo{volume}{238}}, \bibinfo{pages}{471} (\bibinfo{year}{1980}).

\bibitem[{\citenamefont{{Spergel} et~al.}(2006)\citenamefont{{Spergel}, {Bean},
  {Dore'}, {Nolta}, {Bennett}, {Hinshaw}, {Jarosik}, {Komatsu}, {Page},
  {Peiris} et~al.}}]{WMAP3}
\bibinfo{author}{\bibfnamefont{D.~N.} \bibnamefont{{Spergel}}},
  \bibinfo{author}{\bibfnamefont{R.}~\bibnamefont{{Bean}}},
  \bibinfo{author}{\bibfnamefont{O.}~\bibnamefont{{Dore'}}},
  \bibinfo{author}{\bibfnamefont{M.~R.} \bibnamefont{{Nolta}}},
  \bibinfo{author}{\bibfnamefont{C.~L.} \bibnamefont{{Bennett}}},
  \bibinfo{author}{\bibfnamefont{G.}~\bibnamefont{{Hinshaw}}},
  \bibinfo{author}{\bibfnamefont{N.}~\bibnamefont{{Jarosik}}},
  \bibinfo{author}{\bibfnamefont{E.}~\bibnamefont{{Komatsu}}},
  \bibinfo{author}{\bibfnamefont{L.}~\bibnamefont{{Page}}},
  \bibinfo{author}{\bibfnamefont{H.~V.} \bibnamefont{{Peiris}}},
  \bibnamefont{et~al.}, \bibinfo{journal}{ArXiv Astrophysics e-prints}
  (\bibinfo{year}{2006}), \eprint{astro-ph/0603449}.

\bibitem[{\citenamefont{{Bahcall} et~al.}(1999)\citenamefont{{Bahcall},
  {Ostriker}, {Perlmutter}, and {Steinhardt}}}]{Bahcall}
\bibinfo{author}{\bibfnamefont{N.~A.} \bibnamefont{{Bahcall}}},
  \bibinfo{author}{\bibfnamefont{J.~P.} \bibnamefont{{Ostriker}}},
  \bibinfo{author}{\bibfnamefont{S.}~\bibnamefont{{Perlmutter}}},
  \bibnamefont{and} \bibinfo{author}{\bibfnamefont{P.~J.}
  \bibnamefont{{Steinhardt}}}, \bibinfo{journal}{Science}
  \textbf{\bibinfo{volume}{284}}, \bibinfo{pages}{1481} (\bibinfo{year}{1999}).

\bibitem[{\citenamefont{{Navarro} et~al.}(1996)\citenamefont{{Navarro},
  {Frenk}, and {White}}}]{NFW}
\bibinfo{author}{\bibfnamefont{J.~F.} \bibnamefont{{Navarro}}},
  \bibinfo{author}{\bibfnamefont{C.~S.} \bibnamefont{{Frenk}}},
  \bibnamefont{and} \bibinfo{author}{\bibfnamefont{S.~D.~M.}
  \bibnamefont{{White}}}, \bibinfo{journal}{The Astrophysical Journal}
  \textbf{\bibinfo{volume}{462}}, \bibinfo{pages}{563} (\bibinfo{year}{1996}).

\bibitem[{\citenamefont{{Kravtsov} et~al.}(1998)\citenamefont{{Kravtsov},
  {Klypin}, {Bullock}, and {Primack}}}]{Kravtsov}
\bibinfo{author}{\bibfnamefont{A.~V.} \bibnamefont{{Kravtsov}}},
  \bibinfo{author}{\bibfnamefont{A.~A.} \bibnamefont{{Klypin}}},
  \bibinfo{author}{\bibfnamefont{J.~S.} \bibnamefont{{Bullock}}},
  \bibnamefont{and} \bibinfo{author}{\bibfnamefont{J.~R.}
  \bibnamefont{{Primack}}}, \bibinfo{journal}{The Astrophysical Journal}
  \textbf{\bibinfo{volume}{502}}, \bibinfo{pages}{48} (\bibinfo{year}{1998}).

\bibitem[{\citenamefont{{Moore}
  et~al.}(1999{\natexlab{a}})\citenamefont{{Moore}, {Quinn}, {Governato},
  {Stadel}, and {Lake}}}]{Moore}
\bibinfo{author}{\bibfnamefont{B.}~\bibnamefont{{Moore}}},
  \bibinfo{author}{\bibfnamefont{T.}~\bibnamefont{{Quinn}}},
  \bibinfo{author}{\bibfnamefont{F.}~\bibnamefont{{Governato}}},
  \bibinfo{author}{\bibfnamefont{J.}~\bibnamefont{{Stadel}}}, \bibnamefont{and}
  \bibinfo{author}{\bibfnamefont{G.}~\bibnamefont{{Lake}}},
  \bibinfo{journal}{Mon. Not. R. Astron. Soc.} \textbf{\bibinfo{volume}{310}},
  \bibinfo{pages}{1147} (\bibinfo{year}{1999}{\natexlab{a}}).

\bibitem[{\citenamefont{{Ghigna} et~al.}(2000)\citenamefont{{Ghigna}, {Moore},
  {Governato}, {Lake}, {Quinn}, and {Stadel}}}]{Ghigna}
\bibinfo{author}{\bibfnamefont{S.}~\bibnamefont{{Ghigna}}},
  \bibinfo{author}{\bibfnamefont{B.}~\bibnamefont{{Moore}}},
  \bibinfo{author}{\bibfnamefont{F.}~\bibnamefont{{Governato}}},
  \bibinfo{author}{\bibfnamefont{G.}~\bibnamefont{{Lake}}},
  \bibinfo{author}{\bibfnamefont{T.}~\bibnamefont{{Quinn}}}, \bibnamefont{and}
  \bibinfo{author}{\bibfnamefont{J.}~\bibnamefont{{Stadel}}},
  \bibinfo{journal}{The Astrophysical Journal} \textbf{\bibinfo{volume}{544}},
  \bibinfo{pages}{616} (\bibinfo{year}{2000}).

\bibitem[{\citenamefont{{Power} et~al.}(2003)\citenamefont{{Power}, {Navarro},
  {Jenkins}, {Frenk}, {White}, {Springel}, {Stadel}, and {Quinn}}}]{Power03}
\bibinfo{author}{\bibfnamefont{C.}~\bibnamefont{{Power}}},
  \bibinfo{author}{\bibfnamefont{J.~F.} \bibnamefont{{Navarro}}},
  \bibinfo{author}{\bibfnamefont{A.}~\bibnamefont{{Jenkins}}},
  \bibinfo{author}{\bibfnamefont{C.~S.} \bibnamefont{{Frenk}}},
  \bibinfo{author}{\bibfnamefont{S.~D.~M.} \bibnamefont{{White}}},
  \bibinfo{author}{\bibfnamefont{V.}~\bibnamefont{{Springel}}},
  \bibinfo{author}{\bibfnamefont{J.}~\bibnamefont{{Stadel}}}, \bibnamefont{and}
  \bibinfo{author}{\bibfnamefont{T.}~\bibnamefont{{Quinn}}},
  \bibinfo{journal}{Mon. Not. R. Astron. Soc.} \textbf{\bibinfo{volume}{338}},
  \bibinfo{pages}{14} (\bibinfo{year}{2003}), \eprint{astro-ph/0201544}.

\bibitem[{\citenamefont{{Navarro} et~al.}(2004)\citenamefont{{Navarro},
  {Hayashi}, {Power}, {Jenkins}, {Frenk}, {White}, {Springel}, {Stadel}, and
  {Quinn}}}]{Navarro04}
\bibinfo{author}{\bibfnamefont{J.~F.} \bibnamefont{{Navarro}}},
  \bibinfo{author}{\bibfnamefont{E.}~\bibnamefont{{Hayashi}}},
  \bibinfo{author}{\bibfnamefont{C.}~\bibnamefont{{Power}}},
  \bibinfo{author}{\bibfnamefont{A.~R.} \bibnamefont{{Jenkins}}},
  \bibinfo{author}{\bibfnamefont{C.~S.} \bibnamefont{{Frenk}}},
  \bibinfo{author}{\bibfnamefont{S.~D.~M.} \bibnamefont{{White}}},
  \bibinfo{author}{\bibfnamefont{V.}~\bibnamefont{{Springel}}},
  \bibinfo{author}{\bibfnamefont{J.}~\bibnamefont{{Stadel}}}, \bibnamefont{and}
  \bibinfo{author}{\bibfnamefont{T.~R.} \bibnamefont{{Quinn}}},
  \bibinfo{journal}{Mon. Not. R. Astron. Soc.} \textbf{\bibinfo{volume}{349}},
  \bibinfo{pages}{1039} (\bibinfo{year}{2004}), \eprint{astro-ph/0311231}.

\bibitem[{\citenamefont{{Hayashi} et~al.}(2004)\citenamefont{{Hayashi},
  {Navarro}, {Power}, {Jenkins}, {Frenk}, {White}, {Springel}, {Stadel}, and
  {Quinn}}}]{Hayashi04}
\bibinfo{author}{\bibfnamefont{E.}~\bibnamefont{{Hayashi}}},
  \bibinfo{author}{\bibfnamefont{J.~F.} \bibnamefont{{Navarro}}},
  \bibinfo{author}{\bibfnamefont{C.}~\bibnamefont{{Power}}},
  \bibinfo{author}{\bibfnamefont{A.}~\bibnamefont{{Jenkins}}},
  \bibinfo{author}{\bibfnamefont{C.~S.} \bibnamefont{{Frenk}}},
  \bibinfo{author}{\bibfnamefont{S.~D.~M.} \bibnamefont{{White}}},
  \bibinfo{author}{\bibfnamefont{V.}~\bibnamefont{{Springel}}},
  \bibinfo{author}{\bibfnamefont{J.}~\bibnamefont{{Stadel}}}, \bibnamefont{and}
  \bibinfo{author}{\bibfnamefont{T.~R.} \bibnamefont{{Quinn}}},
  \bibinfo{journal}{Mon. Not. R. Astron. Soc.} \textbf{\bibinfo{volume}{355}},
  \bibinfo{pages}{794} (\bibinfo{year}{2004}), \eprint{astro-ph/0310576}.

\bibitem[{\citenamefont{{Diemand} et~al.}(2004)\citenamefont{{Diemand},
  {Moore}, and {Stadel}}}]{Diemand04}
\bibinfo{author}{\bibfnamefont{J.}~\bibnamefont{{Diemand}}},
  \bibinfo{author}{\bibfnamefont{B.}~\bibnamefont{{Moore}}}, \bibnamefont{and}
  \bibinfo{author}{\bibfnamefont{J.}~\bibnamefont{{Stadel}}},
  \bibinfo{journal}{Mon. Not. R. Astron. Soc.} \textbf{\bibinfo{volume}{353}},
  \bibinfo{pages}{624} (\bibinfo{year}{2004}), \eprint{astro-ph/0402267}.

\bibitem[{\citenamefont{{Diemand} et~al.}(2005)\citenamefont{{Diemand}, {Zemp},
  {Moore}, {Stadel}, and {Carollo}}}]{Diemand05}
\bibinfo{author}{\bibfnamefont{J.}~\bibnamefont{{Diemand}}},
  \bibinfo{author}{\bibfnamefont{M.}~\bibnamefont{{Zemp}}},
  \bibinfo{author}{\bibfnamefont{B.}~\bibnamefont{{Moore}}},
  \bibinfo{author}{\bibfnamefont{J.}~\bibnamefont{{Stadel}}}, \bibnamefont{and}
  \bibinfo{author}{\bibfnamefont{C.~M.} \bibnamefont{{Carollo}}},
  \bibinfo{journal}{Mon. Not. R. Astron. Soc.} \textbf{\bibinfo{volume}{364}},
  \bibinfo{pages}{665} (\bibinfo{year}{2005}), \eprint{astro-ph/0504215}.

\bibitem[{\citenamefont{{Tyson} et~al.}(1998)\citenamefont{{Tyson},
  {Kochanski}, and {dell'Antonio}}}]{Tyson}
\bibinfo{author}{\bibfnamefont{J.~A.} \bibnamefont{{Tyson}}},
  \bibinfo{author}{\bibfnamefont{G.~P.} \bibnamefont{{Kochanski}}},
  \bibnamefont{and} \bibinfo{author}{\bibfnamefont{I.~P.}
  \bibnamefont{{dell'Antonio}}}, \bibinfo{journal}{The Astrophysical Journal
  Letters} \textbf{\bibinfo{volume}{498}}, \bibinfo{pages}{L107+}
  (\bibinfo{year}{1998}).

\bibitem[{\citenamefont{{Sand} et~al.}(2004)\citenamefont{{Sand}, {Treu},
  {Smith}, and {Ellis}}}]{Sand04}
\bibinfo{author}{\bibfnamefont{D.~J.} \bibnamefont{{Sand}}},
  \bibinfo{author}{\bibfnamefont{T.}~\bibnamefont{{Treu}}},
  \bibinfo{author}{\bibfnamefont{G.~P.} \bibnamefont{{Smith}}},
  \bibnamefont{and} \bibinfo{author}{\bibfnamefont{R.~S.}
  \bibnamefont{{Ellis}}}, \bibinfo{journal}{The Astrophysical Journal}
  \textbf{\bibinfo{volume}{604}}, \bibinfo{pages}{88} (\bibinfo{year}{2004}),
  \eprint{astro-ph/0310703}.

\bibitem[{\citenamefont{{Katayama} and {Hayashida}}(2004)}]{KH04}
\bibinfo{author}{\bibfnamefont{H.}~\bibnamefont{{Katayama}}} \bibnamefont{and}
  \bibinfo{author}{\bibfnamefont{K.}~\bibnamefont{{Hayashida}}},
  \bibinfo{journal}{Advances in Space Research} \textbf{\bibinfo{volume}{34}},
  \bibinfo{pages}{2519} (\bibinfo{year}{2004}), \eprint{astro-ph/0405363}.

\bibitem[{\citenamefont{{Pointecouteau}
  et~al.}(2005)\citenamefont{{Pointecouteau}, {Arnaud}, and {Pratt}}}]{PAP05}
\bibinfo{author}{\bibfnamefont{E.}~\bibnamefont{{Pointecouteau}}},
  \bibinfo{author}{\bibfnamefont{M.}~\bibnamefont{{Arnaud}}}, \bibnamefont{and}
  \bibinfo{author}{\bibfnamefont{G.~W.} \bibnamefont{{Pratt}}},
  \bibinfo{journal}{Astronomy and Astrophysics} \textbf{\bibinfo{volume}{435}},
  \bibinfo{pages}{1} (\bibinfo{year}{2005}), \eprint{astro-ph/0501635}.

\bibitem[{\citenamefont{{Voigt} and {Fabian}}(2006)}]{VF06}
\bibinfo{author}{\bibfnamefont{L.~M.} \bibnamefont{{Voigt}}} \bibnamefont{and}
  \bibinfo{author}{\bibfnamefont{A.~C.} \bibnamefont{{Fabian}}},
  \bibinfo{journal}{Mon. Not. R. Astron. Soc.} \textbf{\bibinfo{volume}{368}},
  \bibinfo{pages}{518} (\bibinfo{year}{2006}), \eprint{astro-ph/0602373}.

\bibitem[{\citenamefont{{Flores} and {Primack}}(1994)}]{Flores}
\bibinfo{author}{\bibfnamefont{R.~A.} \bibnamefont{{Flores}}} \bibnamefont{and}
  \bibinfo{author}{\bibfnamefont{J.~R.} \bibnamefont{{Primack}}},
  \bibinfo{journal}{The Astrophysical Journal Letters}
  \textbf{\bibinfo{volume}{427}}, \bibinfo{pages}{L1} (\bibinfo{year}{1994}).

\bibitem[{\citenamefont{{Moore}}(1994)}]{Moore94}
\bibinfo{author}{\bibfnamefont{B.}~\bibnamefont{{Moore}}},
  \bibinfo{journal}{Nature} \textbf{\bibinfo{volume}{370}},
  \bibinfo{pages}{629} (\bibinfo{year}{1994}).

\bibitem[{\citenamefont{{Burkert}}(1995)}]{Burk}
\bibinfo{author}{\bibfnamefont{A.}~\bibnamefont{{Burkert}}},
  \bibinfo{journal}{The Astrophysical Journal Letters}
  \textbf{\bibinfo{volume}{447}}, \bibinfo{pages}{L25+} (\bibinfo{year}{1995}).

\bibitem[{\citenamefont{{de Blok} and {McGaugh}}(1997)}]{deBlok}
\bibinfo{author}{\bibfnamefont{W.~J.~G.} \bibnamefont{{de Blok}}}
  \bibnamefont{and} \bibinfo{author}{\bibfnamefont{S.~S.}
  \bibnamefont{{McGaugh}}}, \bibinfo{journal}{Mon. Not. R. Astron. Soc.}
  \textbf{\bibinfo{volume}{290}}, \bibinfo{pages}{533} (\bibinfo{year}{1997}).

\bibitem[{\citenamefont{{McGaugh} and {de Blok}}(1998)}]{MB98}
\bibinfo{author}{\bibfnamefont{S.~S.} \bibnamefont{{McGaugh}}}
  \bibnamefont{and} \bibinfo{author}{\bibfnamefont{W.~J.~G.} \bibnamefont{{de
  Blok}}}, \bibinfo{journal}{The Astrophysical Journal}
  \textbf{\bibinfo{volume}{499}}, \bibinfo{pages}{41} (\bibinfo{year}{1998}),
  \eprint{astro-ph/9801123}.

\bibitem[{\citenamefont{{de Blok} et~al.}(2001)\citenamefont{{de Blok},
  {McGaugh}, and {Rubin}}}]{BMR01}
\bibinfo{author}{\bibfnamefont{W.~J.~G.} \bibnamefont{{de Blok}}},
  \bibinfo{author}{\bibfnamefont{S.~S.} \bibnamefont{{McGaugh}}},
  \bibnamefont{and} \bibinfo{author}{\bibfnamefont{V.~C.}
  \bibnamefont{{Rubin}}}, \bibinfo{journal}{The Astronomical Journal}
  \textbf{\bibinfo{volume}{122}}, \bibinfo{pages}{2396} (\bibinfo{year}{2001}).

\bibitem[{\citenamefont{{Marchesini} et~al.}(2002)\citenamefont{{Marchesini},
  {D'Onghia}, {Chincarini}, {Firmani}, {Conconi}, {Molinari}, and
  {Zacchei}}}]{Marchesini02}
\bibinfo{author}{\bibfnamefont{D.}~\bibnamefont{{Marchesini}}},
  \bibinfo{author}{\bibfnamefont{E.}~\bibnamefont{{D'Onghia}}},
  \bibinfo{author}{\bibfnamefont{G.}~\bibnamefont{{Chincarini}}},
  \bibinfo{author}{\bibfnamefont{C.}~\bibnamefont{{Firmani}}},
  \bibinfo{author}{\bibfnamefont{P.}~\bibnamefont{{Conconi}}},
  \bibinfo{author}{\bibfnamefont{E.}~\bibnamefont{{Molinari}}},
  \bibnamefont{and}
  \bibinfo{author}{\bibfnamefont{A.}~\bibnamefont{{Zacchei}}},
  \bibinfo{journal}{The Astrophysical Journal} \textbf{\bibinfo{volume}{575}},
  \bibinfo{pages}{801} (\bibinfo{year}{2002}), \eprint{astro-ph/0202075}.

\bibitem[{\citenamefont{{Binney} and {Evans}}(2001)}]{BE01}
\bibinfo{author}{\bibfnamefont{J.~J.} \bibnamefont{{Binney}}} \bibnamefont{and}
  \bibinfo{author}{\bibfnamefont{N.~W.} \bibnamefont{{Evans}}},
  \bibinfo{journal}{Mon. Not. R. Astron. Soc.} \textbf{\bibinfo{volume}{327}},
  \bibinfo{pages}{L27} (\bibinfo{year}{2001}), \eprint{astro-ph/0108505}.

\bibitem[{\citenamefont{{Salucci}}(2001)}]{Salucci01}
\bibinfo{author}{\bibfnamefont{P.}~\bibnamefont{{Salucci}}},
  \bibinfo{journal}{Mon. Not. R. Astron. Soc.} \textbf{\bibinfo{volume}{320}},
  \bibinfo{pages}{L1} (\bibinfo{year}{2001}), \eprint{astro-ph/0007389}.

\bibitem[{\citenamefont{{Simon} et~al.}(2005)\citenamefont{{Simon}, {Bolatto},
  {Leroy}, {Blitz}, and {Gates}}}]{SBLBG05}
\bibinfo{author}{\bibfnamefont{J.~D.} \bibnamefont{{Simon}}},
  \bibinfo{author}{\bibfnamefont{A.~D.} \bibnamefont{{Bolatto}}},
  \bibinfo{author}{\bibfnamefont{A.}~\bibnamefont{{Leroy}}},
  \bibinfo{author}{\bibfnamefont{L.}~\bibnamefont{{Blitz}}}, \bibnamefont{and}
  \bibinfo{author}{\bibfnamefont{E.~L.} \bibnamefont{{Gates}}},
  \bibinfo{journal}{The Astrophysical Journal} \textbf{\bibinfo{volume}{621}},
  \bibinfo{pages}{757} (\bibinfo{year}{2005}), \eprint{astro-ph/0412035}.

\bibitem[{\citenamefont{{Moore}
  et~al.}(1999{\natexlab{b}})\citenamefont{{Moore}, {Ghigna}, {Governato},
  {Lake}, {Quinn}, {Stadel}, and {Tozzi}}}]{GSSMoore}
\bibinfo{author}{\bibfnamefont{B.}~\bibnamefont{{Moore}}},
  \bibinfo{author}{\bibfnamefont{S.}~\bibnamefont{{Ghigna}}},
  \bibinfo{author}{\bibfnamefont{F.}~\bibnamefont{{Governato}}},
  \bibinfo{author}{\bibfnamefont{G.}~\bibnamefont{{Lake}}},
  \bibinfo{author}{\bibfnamefont{T.}~\bibnamefont{{Quinn}}},
  \bibinfo{author}{\bibfnamefont{J.}~\bibnamefont{{Stadel}}}, \bibnamefont{and}
  \bibinfo{author}{\bibfnamefont{P.}~\bibnamefont{{Tozzi}}},
  \bibinfo{journal}{The Astrophysical Journal Letters}
  \textbf{\bibinfo{volume}{524}}, \bibinfo{pages}{L19}
  (\bibinfo{year}{1999}{\natexlab{b}}).

\bibitem[{\citenamefont{{Klypin} et~al.}(1999)\citenamefont{{Klypin},
  {Kravtsov}, {Valenzuela}, and {Prada}}}]{KKVP99}
\bibinfo{author}{\bibfnamefont{A.}~\bibnamefont{{Klypin}}},
  \bibinfo{author}{\bibfnamefont{A.~V.} \bibnamefont{{Kravtsov}}},
  \bibinfo{author}{\bibfnamefont{O.}~\bibnamefont{{Valenzuela}}},
  \bibnamefont{and} \bibinfo{author}{\bibfnamefont{F.}~\bibnamefont{{Prada}}},
  \bibinfo{journal}{The Astrophysical Journal} \textbf{\bibinfo{volume}{522}},
  \bibinfo{pages}{82} (\bibinfo{year}{1999}), \eprint{astro-ph/9901240}.

\bibitem[{\citenamefont{{D'Onghia} and {Lake}}(2004)}]{DL04}
\bibinfo{author}{\bibfnamefont{E.}~\bibnamefont{{D'Onghia}}} \bibnamefont{and}
  \bibinfo{author}{\bibfnamefont{G.}~\bibnamefont{{Lake}}},
  \bibinfo{journal}{The Astrophysical Journal} \textbf{\bibinfo{volume}{612}},
  \bibinfo{pages}{628} (\bibinfo{year}{2004}), \eprint{astro-ph/0309735}.

\bibitem[{\citenamefont{{El-Zant} et~al.}(2004)\citenamefont{{El-Zant},
  {Hoffman}, {Primack}, {Combes}, and {Shlosman}}}]{ZHPC04}
\bibinfo{author}{\bibfnamefont{A.~A.} \bibnamefont{{El-Zant}}},
  \bibinfo{author}{\bibfnamefont{Y.}~\bibnamefont{{Hoffman}}},
  \bibinfo{author}{\bibfnamefont{J.}~\bibnamefont{{Primack}}},
  \bibinfo{author}{\bibfnamefont{F.}~\bibnamefont{{Combes}}}, \bibnamefont{and}
  \bibinfo{author}{\bibfnamefont{I.}~\bibnamefont{{Shlosman}}},
  \bibinfo{journal}{The Astrophysical Journal Letters}
  \textbf{\bibinfo{volume}{607}}, \bibinfo{pages}{L75} (\bibinfo{year}{2004}),
  \eprint{astro-ph/0309412}.

\bibitem[{\citenamefont{{Hayashi} and {Navarro}}(2006)}]{HN06}
\bibinfo{author}{\bibfnamefont{E.}~\bibnamefont{{Hayashi}}} \bibnamefont{and}
  \bibinfo{author}{\bibfnamefont{J.~F.} \bibnamefont{{Navarro}}},
  \bibinfo{journal}{Mon. Not. R. Astron. Soc.} \textbf{\bibinfo{volume}{373}},
  \bibinfo{pages}{1117} (\bibinfo{year}{2006}), \eprint{astro-ph/0608376}.

\bibitem[{\citenamefont{{Bullock} et~al.}(2000)\citenamefont{{Bullock},
  {Kravtsov}, and {Weinberg}}}]{BKW00}
\bibinfo{author}{\bibfnamefont{J.~S.} \bibnamefont{{Bullock}}},
  \bibinfo{author}{\bibfnamefont{A.~V.} \bibnamefont{{Kravtsov}}},
  \bibnamefont{and} \bibinfo{author}{\bibfnamefont{D.~H.}
  \bibnamefont{{Weinberg}}}, \bibinfo{journal}{The Astrophysical Journal}
  \textbf{\bibinfo{volume}{539}}, \bibinfo{pages}{517} (\bibinfo{year}{2000}),
  \eprint{astro-ph/0002214}.

\bibitem[{\citenamefont{{Benson} et~al.}(2002)\citenamefont{{Benson}, {Frenk},
  {Lacey}, {Baugh}, and {Cole}}}]{Benson02}
\bibinfo{author}{\bibfnamefont{A.~J.} \bibnamefont{{Benson}}},
  \bibinfo{author}{\bibfnamefont{C.~S.} \bibnamefont{{Frenk}}},
  \bibinfo{author}{\bibfnamefont{C.~G.} \bibnamefont{{Lacey}}},
  \bibinfo{author}{\bibfnamefont{C.~M.} \bibnamefont{{Baugh}}},
  \bibnamefont{and} \bibinfo{author}{\bibfnamefont{S.}~\bibnamefont{{Cole}}},
  \bibinfo{journal}{Mon. Not. R. Astron. Soc.} \textbf{\bibinfo{volume}{333}},
  \bibinfo{pages}{177} (\bibinfo{year}{2002}), \eprint{astro-ph/0108218}.

\bibitem[{\citenamefont{{Kravtsov} et~al.}(2004)\citenamefont{{Kravtsov},
  {Gnedin}, and {Klypin}}}]{KGK04}
\bibinfo{author}{\bibfnamefont{A.~V.} \bibnamefont{{Kravtsov}}},
  \bibinfo{author}{\bibfnamefont{O.~Y.} \bibnamefont{{Gnedin}}},
  \bibnamefont{and} \bibinfo{author}{\bibfnamefont{A.~A.}
  \bibnamefont{{Klypin}}}, \bibinfo{journal}{The Astrophysical Journal}
  \textbf{\bibinfo{volume}{609}}, \bibinfo{pages}{482} (\bibinfo{year}{2004}),
  \eprint{astro-ph/0401088}.

\bibitem[{\citenamefont{{Moore} et~al.}(2006)\citenamefont{{Moore}, {Diemand},
  {Madau}, {Zemp}, and {Stadel}}}]{Moore06}
\bibinfo{author}{\bibfnamefont{B.}~\bibnamefont{{Moore}}},
  \bibinfo{author}{\bibfnamefont{J.}~\bibnamefont{{Diemand}}},
  \bibinfo{author}{\bibfnamefont{P.}~\bibnamefont{{Madau}}},
  \bibinfo{author}{\bibfnamefont{M.}~\bibnamefont{{Zemp}}}, \bibnamefont{and}
  \bibinfo{author}{\bibfnamefont{J.}~\bibnamefont{{Stadel}}},
  \bibinfo{journal}{Mon. Not. R. Astron. Soc.} \textbf{\bibinfo{volume}{368}},
  \bibinfo{pages}{563} (\bibinfo{year}{2006}), \eprint{astro-ph/0510370}.

\bibitem[{\citenamefont{{Spergel} and {Steinhardt}}(2000)}]{oriSIDM}
\bibinfo{author}{\bibfnamefont{D.~N.} \bibnamefont{{Spergel}}}
  \bibnamefont{and} \bibinfo{author}{\bibfnamefont{P.~J.}
  \bibnamefont{{Steinhardt}}}, \bibinfo{journal}{Physical Review Letters}
  \textbf{\bibinfo{volume}{84}}, \bibinfo{pages}{3760} (\bibinfo{year}{2000}).

\bibitem[{\citenamefont{{Wandelt} et~al.}(2001)\citenamefont{{Wandelt}, {Dav{\'
  e}}, {Farrar}, {McGuire}, {Spergel}, and {Steinhardt}}}]{Wandelt}
\bibinfo{author}{\bibfnamefont{B.~D.} \bibnamefont{{Wandelt}}},
  \bibinfo{author}{\bibfnamefont{R.}~\bibnamefont{{Dav{\' e}}}},
  \bibinfo{author}{\bibfnamefont{G.~R.} \bibnamefont{{Farrar}}},
  \bibinfo{author}{\bibfnamefont{P.~C.} \bibnamefont{{McGuire}}},
  \bibinfo{author}{\bibfnamefont{D.~N.} \bibnamefont{{Spergel}}},
  \bibnamefont{and} \bibinfo{author}{\bibfnamefont{P.~J.}
  \bibnamefont{{Steinhardt}}}, in \emph{\bibinfo{booktitle}{Sources and
  Detection of Dark Matter and Dark Energy in the Universe}}, edited by
  \bibinfo{editor}{\bibfnamefont{D.~B.} \bibnamefont{{Cline}}}
  (\bibinfo{publisher}{Springer-Verlag, Berlin, New York},
  \bibinfo{year}{2001}), p. \bibinfo{pages}{263}, \eprint{astro-ph/0006344}.

\bibitem[{\citenamefont{{Dav{\' e}} et~al.}(2001)\citenamefont{{Dav{\' e}},
  {Spergel}, {Steinhardt}, and {Wandelt}}}]{Dave}
\bibinfo{author}{\bibfnamefont{R.}~\bibnamefont{{Dav{\' e}}}},
  \bibinfo{author}{\bibfnamefont{D.~N.} \bibnamefont{{Spergel}}},
  \bibinfo{author}{\bibfnamefont{P.~J.} \bibnamefont{{Steinhardt}}},
  \bibnamefont{and} \bibinfo{author}{\bibfnamefont{B.~D.}
  \bibnamefont{{Wandelt}}}, \bibinfo{journal}{The Astrophysical Journal}
  \textbf{\bibinfo{volume}{547}}, \bibinfo{pages}{574} (\bibinfo{year}{2001}).

\bibitem[{\citenamefont{{Ahn} and {Shapiro}}(2005)}]{AS05}
\bibinfo{author}{\bibfnamefont{K.}~\bibnamefont{{Ahn}}} \bibnamefont{and}
  \bibinfo{author}{\bibfnamefont{P.~R.} \bibnamefont{{Shapiro}}},
  \bibinfo{journal}{Mon. Not. R. Astron. Soc.} \textbf{\bibinfo{volume}{363}},
  \bibinfo{pages}{1092} (\bibinfo{year}{2005}), \eprint{astro-ph/0412169}.

\bibitem[{\citenamefont{Farrar}(2003)}]{Farrar03}
\bibinfo{author}{\bibfnamefont{G.~R.} \bibnamefont{Farrar}},
  \bibinfo{journal}{Int. J. Theor. Phys.} \textbf{\bibinfo{volume}{42}},
  \bibinfo{pages}{1211} (\bibinfo{year}{2003}).

\bibitem[{\citenamefont{{Farrar} and {Zaharijas}}(2006)}]{FZ06}
\bibinfo{author}{\bibfnamefont{G.~R.} \bibnamefont{{Farrar}}} \bibnamefont{and}
  \bibinfo{author}{\bibfnamefont{G.}~\bibnamefont{{Zaharijas}}},
  \bibinfo{journal}{Physical Review Letters} \textbf{\bibinfo{volume}{96}},
  \bibinfo{pages}{041302} (\bibinfo{year}{2006}), \eprint{hep-ph/0510079}.

\bibitem[{\citenamefont{{Kusenko} and {Steinhardt}}(2001)}]{KS01}
\bibinfo{author}{\bibfnamefont{A.}~\bibnamefont{{Kusenko}}} \bibnamefont{and}
  \bibinfo{author}{\bibfnamefont{P.~J.} \bibnamefont{{Steinhardt}}},
  \bibinfo{journal}{Physical Review Letters} \textbf{\bibinfo{volume}{87}},
  \bibinfo{pages}{141301} (\bibinfo{year}{2001}), \eprint{astro-ph/0106008}.

\bibitem[{\citenamefont{Khlopov}(2006)}]{Khlopov}
\bibinfo{author}{\bibfnamefont{M.~Y.} \bibnamefont{Khlopov}},
  \bibinfo{journal}{Pisma Zh. Eksp. Teor. Fiz.} \textbf{\bibinfo{volume}{83}},
  \bibinfo{pages}{3} (\bibinfo{year}{2006}), \eprint{astro-ph/0511796}.

\bibitem[{\citenamefont{{Starkman} et~al.}(1990)\citenamefont{{Starkman},
  {Gould}, {Esmailzadeh}, and {Dimopoulos}}}]{Stark}
\bibinfo{author}{\bibfnamefont{G.~D.} \bibnamefont{{Starkman}}},
  \bibinfo{author}{\bibfnamefont{A.}~\bibnamefont{{Gould}}},
  \bibinfo{author}{\bibfnamefont{R.}~\bibnamefont{{Esmailzadeh}}},
  \bibnamefont{and}
  \bibinfo{author}{\bibfnamefont{S.}~\bibnamefont{{Dimopoulos}}},
  \bibinfo{journal}{Physical Review D} \textbf{\bibinfo{volume}{41}},
  \bibinfo{pages}{3594} (\bibinfo{year}{1990}).

\bibitem[{\citenamefont{{McGuire}}(1994)}]{McGuire94}
\bibinfo{author}{\bibfnamefont{P.~C.} \bibnamefont{{McGuire}}}, Ph.D. thesis,
  \bibinfo{school}{University of Arizona} (\bibinfo{year}{1994}).

\bibitem[{\citenamefont{{McGuire} et~al.}(1995)\citenamefont{{McGuire},
  {Bowen}, {Barker}, {Halverson}, {Kendall}, {Metcalfe}, {Norton}, {Pifer},
  {Barbier}, {Christian} et~al.}}]{McGuire95}
\bibinfo{author}{\bibfnamefont{P.~C.} \bibnamefont{{McGuire}}},
  \bibinfo{author}{\bibfnamefont{T.}~\bibnamefont{{Bowen}}},
  \bibinfo{author}{\bibfnamefont{D.~L.} \bibnamefont{{Barker}}},
  \bibinfo{author}{\bibfnamefont{P.~G.} \bibnamefont{{Halverson}}},
  \bibinfo{author}{\bibfnamefont{K.~R.} \bibnamefont{{Kendall}}},
  \bibinfo{author}{\bibfnamefont{T.~S.} \bibnamefont{{Metcalfe}}},
  \bibinfo{author}{\bibfnamefont{R.~S.} \bibnamefont{{Norton}}},
  \bibinfo{author}{\bibfnamefont{A.~E.} \bibnamefont{{Pifer}}},
  \bibinfo{author}{\bibfnamefont{L.~M.} \bibnamefont{{Barbier}}},
  \bibinfo{author}{\bibfnamefont{E.~R.} \bibnamefont{{Christian}}},
  \bibnamefont{et~al.}, in \emph{\bibinfo{booktitle}{AIP Conf. Proc. 336: Dark
  Matter}}, edited by \bibinfo{editor}{\bibfnamefont{S.~S.}
  \bibnamefont{{Holt}}} \bibnamefont{and} \bibinfo{editor}{\bibfnamefont{C.~L.}
  \bibnamefont{{Bennett}}} (\bibinfo{year}{1995}), p.~\bibinfo{pages}{53}.

\bibitem[{\citenamefont{{McGuire} and {Steinhardt}}(2001)}]{MS01}
\bibinfo{author}{\bibfnamefont{P.~C.} \bibnamefont{{McGuire}}}
  \bibnamefont{and} \bibinfo{author}{\bibfnamefont{P.~J.}
  \bibnamefont{{Steinhardt}}}, in \emph{\bibinfo{booktitle}{Proceedings of the
  27th International Cosmic Ray Conference, Hamburg, Germany}}
  (\bibinfo{year}{2001}), p. \bibinfo{pages}{1566}, \eprint{astro-ph/0105567}.

\bibitem[{\citenamefont{{Cyburt} et~al.}(2002)\citenamefont{{Cyburt}, {Fields},
  {Pavlidou}, and {Wandelt}}}]{CFPW02}
\bibinfo{author}{\bibfnamefont{R.~H.} \bibnamefont{{Cyburt}}},
  \bibinfo{author}{\bibfnamefont{B.~D.} \bibnamefont{{Fields}}},
  \bibinfo{author}{\bibfnamefont{V.}~\bibnamefont{{Pavlidou}}},
  \bibnamefont{and}
  \bibinfo{author}{\bibfnamefont{B.}~\bibnamefont{{Wandelt}}},
  \bibinfo{journal}{Physical Review D} \textbf{\bibinfo{volume}{65}},
  \bibinfo{pages}{123503} (\bibinfo{year}{2002}), \eprint{astro-ph/0203240}.

\bibitem[{\citenamefont{{Chen} et~al.}(2002)\citenamefont{{Chen}, {Hannestad},
  and {Scherrer}}}]{CHS02}
\bibinfo{author}{\bibfnamefont{X.}~\bibnamefont{{Chen}}},
  \bibinfo{author}{\bibfnamefont{S.}~\bibnamefont{{Hannestad}}},
  \bibnamefont{and} \bibinfo{author}{\bibfnamefont{R.~J.}
  \bibnamefont{{Scherrer}}}, \bibinfo{journal}{Physical Review D}
  \textbf{\bibinfo{volume}{65}}, \bibinfo{pages}{123515}
  (\bibinfo{year}{2002}), \eprint{astro-ph/0202496}.

\bibitem[{\citenamefont{{McCammon} et~al.}(2002)\citenamefont{{McCammon},
  {Almy}, {Apodaca}, {Bergmann Tiest}, {Cui}, {Deiker}, {Galeazzi}, {Juda},
  {Lesser}, {Mihara} et~al.}}]{big}
\bibinfo{author}{\bibfnamefont{D.}~\bibnamefont{{McCammon}}},
  \bibinfo{author}{\bibfnamefont{R.}~\bibnamefont{{Almy}}},
  \bibinfo{author}{\bibfnamefont{E.}~\bibnamefont{{Apodaca}}},
  \bibinfo{author}{\bibfnamefont{W.}~\bibnamefont{{Bergmann Tiest}}},
  \bibinfo{author}{\bibfnamefont{W.}~\bibnamefont{{Cui}}},
  \bibinfo{author}{\bibfnamefont{S.}~\bibnamefont{{Deiker}}},
  \bibinfo{author}{\bibfnamefont{M.}~\bibnamefont{{Galeazzi}}},
  \bibinfo{author}{\bibfnamefont{M.}~\bibnamefont{{Juda}}},
  \bibinfo{author}{\bibfnamefont{A.}~\bibnamefont{{Lesser}}},
  \bibinfo{author}{\bibfnamefont{T.}~\bibnamefont{{Mihara}}},
  \bibnamefont{et~al.}, \bibinfo{journal}{The Astrophysical Journal}
  \textbf{\bibinfo{volume}{576}}, \bibinfo{pages}{188} (\bibinfo{year}{2002}).

\bibitem[{\citenamefont{{Zaharijas} and {Farrar}}(2005)}]{ZF05}
\bibinfo{author}{\bibfnamefont{G.}~\bibnamefont{{Zaharijas}}} \bibnamefont{and}
  \bibinfo{author}{\bibfnamefont{G.~R.} \bibnamefont{{Farrar}}},
  \bibinfo{journal}{Physical Review D} \textbf{\bibinfo{volume}{72}},
  \bibinfo{pages}{083502} (\bibinfo{year}{2005}), \eprint{astro-ph/0406531}.

\bibitem[{\citenamefont{{McCammon} et~al.}(1996)\citenamefont{{McCammon},
  {Almy}, {Deiker}, {Morgenthaler}, {Kelley}, {Marshall}, {Moseley}, {Stahle},
  and {Szymkowiak}}}]{rocket}
\bibinfo{author}{\bibfnamefont{D.}~\bibnamefont{{McCammon}}},
  \bibinfo{author}{\bibfnamefont{R.}~\bibnamefont{{Almy}}},
  \bibinfo{author}{\bibfnamefont{S.}~\bibnamefont{{Deiker}}},
  \bibinfo{author}{\bibfnamefont{J.}~\bibnamefont{{Morgenthaler}}},
  \bibinfo{author}{\bibfnamefont{R.~L.} \bibnamefont{{Kelley}}},
  \bibinfo{author}{\bibfnamefont{F.~J.} \bibnamefont{{Marshall}}},
  \bibinfo{author}{\bibfnamefont{S.~H.} \bibnamefont{{Moseley}}},
  \bibinfo{author}{\bibfnamefont{C.~K.} \bibnamefont{{Stahle}}},
  \bibnamefont{and} \bibinfo{author}{\bibfnamefont{A.~E.}
  \bibnamefont{{Szymkowiak}}}, \bibinfo{journal}{Nuclear Instruments and
  Methods in Physics Research A} \textbf{\bibinfo{volume}{370}},
  \bibinfo{pages}{266} (\bibinfo{year}{1996}).

\bibitem[{\citenamefont{{Stahle} et~al.}(1996)\citenamefont{{Stahle}, {Kelley},
  {McCammon}, {Moseley}, and {Szymkowiak}}}]{microcal}
\bibinfo{author}{\bibfnamefont{C.~K.} \bibnamefont{{Stahle}}},
  \bibinfo{author}{\bibfnamefont{R.~L.} \bibnamefont{{Kelley}}},
  \bibinfo{author}{\bibfnamefont{D.}~\bibnamefont{{McCammon}}},
  \bibinfo{author}{\bibfnamefont{S.~H.} \bibnamefont{{Moseley}}},
  \bibnamefont{and} \bibinfo{author}{\bibfnamefont{A.~E.}
  \bibnamefont{{Szymkowiak}}}, \bibinfo{journal}{Nuclear Instruments and
  Methods in Physics Research A} \textbf{\bibinfo{volume}{370}},
  \bibinfo{pages}{173} (\bibinfo{year}{1996}).

\bibitem[{\citenamefont{{Gates} et~al.}(1995)\citenamefont{{Gates}, {Gyuk}, and
  {Turner}}}]{Gates}
\bibinfo{author}{\bibfnamefont{E.~I.} \bibnamefont{{Gates}}},
  \bibinfo{author}{\bibfnamefont{G.}~\bibnamefont{{Gyuk}}}, \bibnamefont{and}
  \bibinfo{author}{\bibfnamefont{M.~S.} \bibnamefont{{Turner}}},
  \bibinfo{journal}{The Astrophysical Journal Letters}
  \textbf{\bibinfo{volume}{449}}, \bibinfo{pages}{L123+}
  (\bibinfo{year}{1995}).

\bibitem[{\citenamefont{{Moore} et~al.}(2001)\citenamefont{{Moore},
  {Calc{\'a}neo-Rold{\'a}n}, {Stadel}, {Quinn}, {Lake}, {Ghigna}, and
  {Governato}}}]{Moore01}
\bibinfo{author}{\bibfnamefont{B.}~\bibnamefont{{Moore}}},
  \bibinfo{author}{\bibfnamefont{C.}~\bibnamefont{{Calc{\'a}neo-Rold{\'a}n}}},
  \bibinfo{author}{\bibfnamefont{J.}~\bibnamefont{{Stadel}}},
  \bibinfo{author}{\bibfnamefont{T.}~\bibnamefont{{Quinn}}},
  \bibinfo{author}{\bibfnamefont{G.}~\bibnamefont{{Lake}}},
  \bibinfo{author}{\bibfnamefont{S.}~\bibnamefont{{Ghigna}}}, \bibnamefont{and}
  \bibinfo{author}{\bibfnamefont{F.}~\bibnamefont{{Governato}}},
  \bibinfo{journal}{\prd} \textbf{\bibinfo{volume}{64}},
  \bibinfo{pages}{063508} (\bibinfo{year}{2001}), \eprint{astro-ph/0106271}.

\bibitem[{\citenamefont{{Lewin} and {Smith}}(1996)}]{Lewin}
\bibinfo{author}{\bibfnamefont{J.~D.} \bibnamefont{{Lewin}}} \bibnamefont{and}
  \bibinfo{author}{\bibfnamefont{P.~F.} \bibnamefont{{Smith}}},
  \bibinfo{journal}{Astroparticle Physics} \textbf{\bibinfo{volume}{6}},
  \bibinfo{pages}{87} (\bibinfo{year}{1996}).

\bibitem[{\citenamefont{{Green}}(2003)}]{Green03}
\bibinfo{author}{\bibfnamefont{A.~M.} \bibnamefont{{Green}}},
  \bibinfo{journal}{\prd} \textbf{\bibinfo{volume}{68}},
  \bibinfo{pages}{023004} (\bibinfo{year}{2003}), \eprint{astro-ph/0304446}.

\bibitem[{\citenamefont{{Drukier} et~al.}(1986)\citenamefont{{Drukier},
  {Freese}, and {Spergel}}}]{Drukier}
\bibinfo{author}{\bibfnamefont{A.~K.} \bibnamefont{{Drukier}}},
  \bibinfo{author}{\bibfnamefont{K.}~\bibnamefont{{Freese}}}, \bibnamefont{and}
  \bibinfo{author}{\bibfnamefont{D.~N.} \bibnamefont{{Spergel}}},
  \bibinfo{journal}{Physical Review D} \textbf{\bibinfo{volume}{33}},
  \bibinfo{pages}{3495} (\bibinfo{year}{1986}).

\bibitem[{\citenamefont{{Kerr} and {Lynden-Bell}}(1986)}]{v1}
\bibinfo{author}{\bibfnamefont{F.~J.} \bibnamefont{{Kerr}}} \bibnamefont{and}
  \bibinfo{author}{\bibfnamefont{D.}~\bibnamefont{{Lynden-Bell}}},
  \bibinfo{journal}{Mon. Not. R. Astron. Soc.} \textbf{\bibinfo{volume}{221}},
  \bibinfo{pages}{1023} (\bibinfo{year}{1986}).

\bibitem[{\citenamefont{{Caldwell} and {Coulson}}(1987)}]{v2}
\bibinfo{author}{\bibfnamefont{J.~A.~R.} \bibnamefont{{Caldwell}}}
  \bibnamefont{and} \bibinfo{author}{\bibfnamefont{I.~M.}
  \bibnamefont{{Coulson}}}, \bibinfo{journal}{The Astronomical Journal}
  \textbf{\bibinfo{volume}{93}}, \bibinfo{pages}{1090} (\bibinfo{year}{1987}).

\bibitem[{\citenamefont{{Olling} and {Merrifield}}(1998)}]{newconstants}
\bibinfo{author}{\bibfnamefont{R.~P.} \bibnamefont{{Olling}}} \bibnamefont{and}
  \bibinfo{author}{\bibfnamefont{M.~R.} \bibnamefont{{Merrifield}}},
  \bibinfo{journal}{Mon. Not. R. Astron. Soc.} \textbf{\bibinfo{volume}{297}},
  \bibinfo{pages}{943} (\bibinfo{year}{1998}).

\bibitem[{\citenamefont{{Kochanek}}(1996)}]{Kochanek96}
\bibinfo{author}{\bibfnamefont{C.~S.} \bibnamefont{{Kochanek}}},
  \bibinfo{journal}{\apj} \textbf{\bibinfo{volume}{457}}, \bibinfo{pages}{228}
  (\bibinfo{year}{1996}), \eprint{astro-ph/9505068}.

\bibitem[{\citenamefont{{Reid} et~al.}(1999)\citenamefont{{Reid}, {Readhead},
  {Vermeulen}, and {Treuhaft}}}]{v3}
\bibinfo{author}{\bibfnamefont{M.~J.} \bibnamefont{{Reid}}},
  \bibinfo{author}{\bibfnamefont{A.~C.~S.} \bibnamefont{{Readhead}}},
  \bibinfo{author}{\bibfnamefont{R.~C.} \bibnamefont{{Vermeulen}}},
  \bibnamefont{and} \bibinfo{author}{\bibfnamefont{R.~N.}
  \bibnamefont{{Treuhaft}}}, \bibinfo{journal}{The Astrophysical Journal}
  \textbf{\bibinfo{volume}{524}}, \bibinfo{pages}{816} (\bibinfo{year}{1999}).

\bibitem[{\citenamefont{{Olling} and {Dehnen}}(2003)}]{v4}
\bibinfo{author}{\bibfnamefont{R.~P.} \bibnamefont{{Olling}}} \bibnamefont{and}
  \bibinfo{author}{\bibfnamefont{W.}~\bibnamefont{{Dehnen}}},
  \bibinfo{journal}{The Astrophysical Journal} \textbf{\bibinfo{volume}{599}},
  \bibinfo{pages}{275} (\bibinfo{year}{2003}), \eprint{arXiv:astro-ph/0301486}.

\bibitem[{\citenamefont{{Cudworth}}(1990)}]{starvel}
\bibinfo{author}{\bibfnamefont{K.~M.} \bibnamefont{{Cudworth}}},
  \bibinfo{journal}{The Astronomical Journal} \textbf{\bibinfo{volume}{99}},
  \bibinfo{pages}{590} (\bibinfo{year}{1990}).

\bibitem[{\citenamefont{{Leonard} and {Tremaine}}(1990)}]{LT90}
\bibinfo{author}{\bibfnamefont{P.~J.~T.} \bibnamefont{{Leonard}}}
  \bibnamefont{and}
  \bibinfo{author}{\bibfnamefont{S.}~\bibnamefont{{Tremaine}}},
  \bibinfo{journal}{\apj} \textbf{\bibinfo{volume}{353}}, \bibinfo{pages}{486}
  (\bibinfo{year}{1990}).

\bibitem[{\citenamefont{{Smith} et~al.}(2007)\citenamefont{{Smith}, {Ruchti},
  {Helmi}, {Wyse}, {Fulbright}, {Freeman}, {Navarro}, {Seabroke}, {Steinmetz},
  {Williams} et~al.}}]{Rave06}
\bibinfo{author}{\bibfnamefont{M.~C.} \bibnamefont{{Smith}}},
  \bibinfo{author}{\bibfnamefont{G.~R.} \bibnamefont{{Ruchti}}},
  \bibinfo{author}{\bibfnamefont{A.}~\bibnamefont{{Helmi}}},
  \bibinfo{author}{\bibfnamefont{R.~F.~G.} \bibnamefont{{Wyse}}},
  \bibinfo{author}{\bibfnamefont{J.~P.} \bibnamefont{{Fulbright}}},
  \bibinfo{author}{\bibfnamefont{K.~C.} \bibnamefont{{Freeman}}},
  \bibinfo{author}{\bibfnamefont{J.~F.} \bibnamefont{{Navarro}}},
  \bibinfo{author}{\bibfnamefont{G.~M.} \bibnamefont{{Seabroke}}},
  \bibinfo{author}{\bibfnamefont{M.}~\bibnamefont{{Steinmetz}}},
  \bibinfo{author}{\bibfnamefont{M.}~\bibnamefont{{Williams}}},
  \bibnamefont{et~al.}, \bibinfo{journal}{Mon. Not. R. Astron. Soc.}
  \textbf{\bibinfo{volume}{379}}, \bibinfo{pages}{755} (\bibinfo{year}{2007}),
  \eprint{arXiv:astro-ph/0611671}.

\bibitem[{\citenamefont{{Reid}}(1993)}]{solarrad}
\bibinfo{author}{\bibfnamefont{M.~J.} \bibnamefont{{Reid}}},
  \bibinfo{journal}{Annual Review of Astronomy and Astrophysics}
  \textbf{\bibinfo{volume}{31}}, \bibinfo{pages}{345} (\bibinfo{year}{1993}).

\bibitem[{\citenamefont{{McNamara} et~al.}(2000)\citenamefont{{McNamara},
  {Madsen}, {Barnes}, and {Ericksen}}}]{solarrad2}
\bibinfo{author}{\bibfnamefont{D.~H.} \bibnamefont{{McNamara}}},
  \bibinfo{author}{\bibfnamefont{J.~B.} \bibnamefont{{Madsen}}},
  \bibinfo{author}{\bibfnamefont{J.}~\bibnamefont{{Barnes}}}, \bibnamefont{and}
  \bibinfo{author}{\bibfnamefont{B.~F.} \bibnamefont{{Ericksen}}},
  \bibinfo{journal}{The Publications of the Astronomical Society of the
  Pacific} \textbf{\bibinfo{volume}{112}}, \bibinfo{pages}{202}
  (\bibinfo{year}{2000}).

\bibitem[{\citenamefont{{Avedisova}}(2005)}]{Avedisova05}
\bibinfo{author}{\bibfnamefont{V.~S.} \bibnamefont{{Avedisova}}},
  \bibinfo{journal}{Astronomy Reports} \textbf{\bibinfo{volume}{49}},
  \bibinfo{pages}{435} (\bibinfo{year}{2005}).

\bibitem[{\citenamefont{{Dehnen} and {Binney}}(1998)}]{DB98}
\bibinfo{author}{\bibfnamefont{W.}~\bibnamefont{{Dehnen}}} \bibnamefont{and}
  \bibinfo{author}{\bibfnamefont{J.~J.} \bibnamefont{{Binney}}},
  \bibinfo{journal}{Mon. Not. R. Astron. Soc.} \textbf{\bibinfo{volume}{298}},
  \bibinfo{pages}{387} (\bibinfo{year}{1998}), \eprint{astro-ph/9710077}.

\bibitem[{\citenamefont{{Goodman} and {Witten}}(1985)}]{GW85}
\bibinfo{author}{\bibfnamefont{M.~W.} \bibnamefont{{Goodman}}}
  \bibnamefont{and} \bibinfo{author}{\bibfnamefont{E.}~\bibnamefont{{Witten}}},
  \bibinfo{journal}{Physical Review D} \textbf{\bibinfo{volume}{31}},
  \bibinfo{pages}{3059} (\bibinfo{year}{1985}).

\bibitem[{\citenamefont{{Gould}}(1987)}]{Gould87}
\bibinfo{author}{\bibfnamefont{A.}~\bibnamefont{{Gould}}},
  \bibinfo{journal}{The Astrophysical Journal} \textbf{\bibinfo{volume}{321}},
  \bibinfo{pages}{571} (\bibinfo{year}{1987}).

\bibitem[{\citenamefont{{Gelmini} et~al.}(2002)\citenamefont{{Gelmini},
  {Kusenko}, and {Nussinov}}}]{GKN02}
\bibinfo{author}{\bibfnamefont{G.}~\bibnamefont{{Gelmini}}},
  \bibinfo{author}{\bibfnamefont{A.}~\bibnamefont{{Kusenko}}},
  \bibnamefont{and}
  \bibinfo{author}{\bibfnamefont{S.}~\bibnamefont{{Nussinov}}},
  \bibinfo{journal}{Physical Review Letters} \textbf{\bibinfo{volume}{89}},
  \bibinfo{pages}{101302} (\bibinfo{year}{2002}), \eprint{hep-ph/0203179}.

\bibitem[{\citenamefont{{Engel}}(1991)}]{Engel91}
\bibinfo{author}{\bibfnamefont{J.}~\bibnamefont{{Engel}}},
  \bibinfo{journal}{Physics Letters B} \textbf{\bibinfo{volume}{264}},
  \bibinfo{pages}{114} (\bibinfo{year}{1991}).

\bibitem[{\citenamefont{{Freedman}}(1974)}]{Freedman74}
\bibinfo{author}{\bibfnamefont{D.~Z.} \bibnamefont{{Freedman}}},
  \bibinfo{journal}{Physical Review D} \textbf{\bibinfo{volume}{9}},
  \bibinfo{pages}{1389} (\bibinfo{year}{1974}).

\bibitem[{\citenamefont{{Helm}}(1956)}]{Helm56}
\bibinfo{author}{\bibfnamefont{R.~H.} \bibnamefont{{Helm}}},
  \bibinfo{journal}{Physical Review} \textbf{\bibinfo{volume}{104}},
  \bibinfo{pages}{1466} (\bibinfo{year}{1956}).

\bibitem[{\citenamefont{{Matsumoto} and {Nishimura}}(1998)}]{rng}
\bibinfo{author}{\bibfnamefont{M.}~\bibnamefont{{Matsumoto}}} \bibnamefont{and}
  \bibinfo{author}{\bibfnamefont{T.}~\bibnamefont{{Nishimura}}},
  \bibinfo{journal}{ACM Transactions on Modeling and Computer Simulation}
  \textbf{\bibinfo{volume}{8}}, \bibinfo{pages}{3} (\bibinfo{year}{1998}).

\bibitem[{\citenamefont{{Simpson} et~al.}(1980)\citenamefont{{Simpson},
  {Bastian}, {Chenette}, {McKibben}, and {Pyle}}}]{Simpson80}
\bibinfo{author}{\bibfnamefont{J.~A.} \bibnamefont{{Simpson}}},
  \bibinfo{author}{\bibfnamefont{T.~S.} \bibnamefont{{Bastian}}},
  \bibinfo{author}{\bibfnamefont{D.~L.} \bibnamefont{{Chenette}}},
  \bibinfo{author}{\bibfnamefont{R.~B.} \bibnamefont{{McKibben}}},
  \bibnamefont{and} \bibinfo{author}{\bibfnamefont{K.~R.}
  \bibnamefont{{Pyle}}}, \bibinfo{journal}{Journal of Geophysical Research}
  \textbf{\bibinfo{volume}{85}}, \bibinfo{pages}{5731} (\bibinfo{year}{1980}).

\bibitem[{\citenamefont{{Shirk} and {Price}}(1978)}]{SP78}
\bibinfo{author}{\bibfnamefont{E.~K.} \bibnamefont{{Shirk}}} \bibnamefont{and}
  \bibinfo{author}{\bibfnamefont{P.~B.} \bibnamefont{{Price}}},
  \bibinfo{journal}{\apj} \textbf{\bibinfo{volume}{220}}, \bibinfo{pages}{719}
  (\bibinfo{year}{1978}).

\bibitem[{\citenamefont{{Mewaldt} et~al.}(2001)\citenamefont{{Mewaldt},
  {Labrador}, {Lopate}, and {McKibben}}}]{IMP01}
\bibinfo{author}{\bibfnamefont{R.~A.} \bibnamefont{{Mewaldt}}},
  \bibinfo{author}{\bibfnamefont{A.~W.} \bibnamefont{{Labrador}}},
  \bibinfo{author}{\bibfnamefont{C.}~\bibnamefont{{Lopate}}}, \bibnamefont{and}
  \bibinfo{author}{\bibfnamefont{R.~B.} \bibnamefont{{McKibben}}}
  (\bibinfo{year}{2001}), \bibinfo{note}{private communication}.

\bibitem[{\citenamefont{{Rich} et~al.}(1987)\citenamefont{{Rich}, {Rocchia},
  and {Spiro}}}]{RRS87}
\bibinfo{author}{\bibfnamefont{J.}~\bibnamefont{{Rich}}},
  \bibinfo{author}{\bibfnamefont{R.}~\bibnamefont{{Rocchia}}},
  \bibnamefont{and} \bibinfo{author}{\bibfnamefont{M.}~\bibnamefont{{Spiro}}},
  \bibinfo{journal}{Physics Letters B} \textbf{\bibinfo{volume}{194}},
  \bibinfo{pages}{173} (\bibinfo{year}{1987}).

\bibitem[{\citenamefont{{Bacci} et~al.}(1994)\citenamefont{{Bacci}, {Belli},
  {Bernabei}, {Dai}, {Ding}, {di Nicolantonio}, {Gaillard}, {Gerbier}, {Kuang},
  {Incicchitti} et~al.}}]{Bacci94}
\bibinfo{author}{\bibfnamefont{C.}~\bibnamefont{{Bacci}}},
  \bibinfo{author}{\bibfnamefont{P.}~\bibnamefont{{Belli}}},
  \bibinfo{author}{\bibfnamefont{R.}~\bibnamefont{{Bernabei}}},
  \bibinfo{author}{\bibfnamefont{C.}~\bibnamefont{{Dai}}},
  \bibinfo{author}{\bibfnamefont{L.}~\bibnamefont{{Ding}}},
  \bibinfo{author}{\bibfnamefont{W.}~\bibnamefont{{di Nicolantonio}}},
  \bibinfo{author}{\bibfnamefont{E.}~\bibnamefont{{Gaillard}}},
  \bibinfo{author}{\bibfnamefont{G.}~\bibnamefont{{Gerbier}}},
  \bibinfo{author}{\bibfnamefont{H.}~\bibnamefont{{Kuang}}},
  \bibinfo{author}{\bibfnamefont{A.}~\bibnamefont{{Incicchitti}}},
  \bibnamefont{et~al.}, \bibinfo{journal}{Astroparticle Physics}
  \textbf{\bibinfo{volume}{2}}, \bibinfo{pages}{13} (\bibinfo{year}{1994}).

\bibitem[{\citenamefont{{Bernabei} et~al.}(1999)\citenamefont{{Bernabei},
  {Belli}, {Cerulli}, {Montecchia}, {Amato}, {Ignesti}, {Incicchitti},
  {Prosperi}, {Dai}, {He} et~al.}}]{Bernabei99}
\bibinfo{author}{\bibfnamefont{R.}~\bibnamefont{{Bernabei}}},
  \bibinfo{author}{\bibfnamefont{P.}~\bibnamefont{{Belli}}},
  \bibinfo{author}{\bibfnamefont{R.}~\bibnamefont{{Cerulli}}},
  \bibinfo{author}{\bibfnamefont{F.}~\bibnamefont{{Montecchia}}},
  \bibinfo{author}{\bibfnamefont{M.}~\bibnamefont{{Amato}}},
  \bibinfo{author}{\bibfnamefont{G.}~\bibnamefont{{Ignesti}}},
  \bibinfo{author}{\bibfnamefont{A.}~\bibnamefont{{Incicchitti}}},
  \bibinfo{author}{\bibfnamefont{D.}~\bibnamefont{{Prosperi}}},
  \bibinfo{author}{\bibfnamefont{C.~J.} \bibnamefont{{Dai}}},
  \bibinfo{author}{\bibfnamefont{H.~L.} \bibnamefont{{He}}},
  \bibnamefont{et~al.}, \bibinfo{journal}{Physical Review Letters}
  \textbf{\bibinfo{volume}{83}}, \bibinfo{pages}{4918} (\bibinfo{year}{1999}).

\bibitem[{\citenamefont{{Albuquerque} and {Baudis}}(2003{\natexlab{a}})}]{AB93}
\bibinfo{author}{\bibfnamefont{I.~F.~M.} \bibnamefont{{Albuquerque}}}
  \bibnamefont{and} \bibinfo{author}{\bibfnamefont{L.}~\bibnamefont{{Baudis}}},
  \bibinfo{journal}{Physical Review Letters} \textbf{\bibinfo{volume}{90}},
  \bibinfo{pages}{221301} (\bibinfo{year}{2003}{\natexlab{a}}),
  \eprint{astro-ph/0301188}.

\bibitem[{\citenamefont{{Albuquerque} and
  {Baudis}}(2003{\natexlab{b}})}]{AB93E}
\bibinfo{author}{\bibfnamefont{I.~F.~M.} \bibnamefont{{Albuquerque}}}
  \bibnamefont{and} \bibinfo{author}{\bibfnamefont{L.}~\bibnamefont{{Baudis}}},
  \bibinfo{journal}{Physical Review Letters} \textbf{\bibinfo{volume}{91}},
  \bibinfo{pages}{229903(E)} (\bibinfo{year}{2003}{\natexlab{b}}).

\end{thebibliography}
\end{document}